\documentclass[a4paper,12pt]{article}
\pdfoutput=1

\usepackage{jheppub}
\usepackage{booktabs}
\allowdisplaybreaks
\usepackage[utf8]{inputenc} 

\def \NP{{\rm NP}}

\def \GeV{{\,\rm GeV}}

\newcommand{\mB}{{\mathcal B}}

\def\nn{\nonumber}

\def\bra#1{\left\langle #1\right|}
\def\ket#1{\left| #1\right\rangle}
\newcommand{\abs}[1]{\left\vert{#1}\right\vert}

\newenvironment{spmatrix}{\Biggl(\begin{matrix}} {\end{matrix}\Biggr)}

\newcommand{\tred}[1]{\textcolor{black}{#1}}
\newcommand{\tblue}[1]{\textcolor{black}{#1}}

\newcommand{\wilson}[2][]{\mathcal{C}^{#1}_{#2}}
\newcommand{\op}[2][]{\mathcal{O}^{#1}_{#2}}
\newcommand{\dd}{{\rm d}}
\newcommand{\im}{{\rm Im}}
\newcommand{\real}{{\rm Re}}

\usepackage{tikz}
\usetikzlibrary{calc}
\usetikzlibrary{trees}
\usetikzlibrary{decorations.pathmorphing}
\usetikzlibrary{decorations.markings}
\usetikzlibrary{arrows.meta}

\usetikzlibrary{positioning}
\tikzset{>=stealth}

\tikzset{
    fermion/.style={draw=teal, postaction={decorate},decoration={markings,mark=at position .55 with {\arrow[draw=teal]{>}}}},
    fermionbar/.style={draw=teal, postaction={decorate},decoration={markings,mark=at position .55 with {\arrow[draw=teal]{<}}}},
    fermionnoarrow/.style={draw=teal},
    scalar/.style={dashed,draw=teal, postaction={decorate},decoration={markings,mark=at position .55 with {\arrow[draw=teal]{>}}}},
    scalarbar/.style={dashed,draw=teal, postaction={decorate},decoration={markings,mark=at position .55 with {\arrow[draw=teal]{<}}}},
    scalarnoarrow/.style={dotted,draw=teal},
    dotline/.style={dotted,draw=gray},
    photon/.style={decorate, draw=teal,decoration={snake,amplitude=4pt, segment length=5pt} },
    gluon/.style={decorate, draw=teal, decoration={coil,amplitude=4pt, segment length=5pt}},
     Z/.style={decorate, draw=teal,decoration={zigzag,amplitude=5pt, segment length=6pt}},
	 W/.style={decorate, draw=teal,decoration={zigzag,amplitude=3pt, segment length=6pt}},
}

\title{\boldmath Angular distribution of the rare decay $\Lambda_b \to \Lambda(\to N \pi) \ell^+\ell^-$}
\author[a]{Han Yan}
\affiliation[a]{School of Physics and Electric Information,\\Huang gang Normal University, Huang gang, Hubei 438000, China}
\emailAdd{yanhan@mails.ccnu.edu.cn}

\abstract{We provide a determination of the complete angular distribution for the four body rare decay $\Lambda_b \to \Lambda(\to N \pi) \ell^+\ell^-$, with unpolarized $\Lambda_b$ baryons and massive leptons, in the operator basis approach which includes the scalar, pseudo-scalar, vector, axial-vector  and tensor operators.  Especially, the contributions of tensor operators have been calculated for the first time in this work. 
Since the lepton mass is retained in our calculations, the lepton flavour universality and the decay mode  $\Lambda_b \to \Lambda(\to N \pi) \tau^+\tau^-$ can be investigated in detail. For comparison with the experiment, we study the numerical results of observables  within  the Standard Model and the $S_1+S_3$ Leptoquark model. Significant deviation can be found between experiment data and the Standard Model predictions. 
The $S_1+S_3$ Leptoquark model can be further explored with the experimental progresses. In addition, we demonstrate  the sensitivity of various angular observables to tensor operators contributions firstly, and find out that the potential New Physics effects of tensor operators can not be ignored in $b\to s\ell^+\ell^-$ transitions. }

\begin{document}
\maketitle
\flushbottom

\section{Introduction}
\label{sec:intro}
In the Standard Model (SM), the flavour-changing-neutral-current (FCNC) transitions are  not allowed at the tree level because of Glashow-Iliopoulos-Maiani (GIM) mechanism~\cite{Glashow:1970gm}. The FCNC transitions in the SM  are CKM and loop suppressed, and therefore sensitive to New Physics (NP)~\cite{Hurth:2010tk,Blake:2016olu}. The rare $b\to s\ell^+\ell^-$ transitions induced by loop process are well known FCNC transitions. 
Although the Large Hadron Collider (LHC) has not observed any new particles beyond the SM directly so far, a series of tensions between the experimental data and the predictions of SM have been found, such as, the ratios $R_{K^{(*)}}\equiv \mB(B \to K^{(*)} \mu^+ \mu^-)/\mB(B \to K^{(*)} e^+ e^-)$ of $B\to K^{(*)}\ell^+\ell^-$~\cite{Aaij:2014ora,Aaij:2017vbb,Bordone:2016gaq}, the differential decay branching fraction of $B_{s}\to \phi \mu^+\mu^-$~\cite{Aaij:2013aln,Horgan:2013pva,Aaij:2014pli,Aaij:2015esa,Bouchard:2013mia}, as well as the angular observables in $B\to K^{*}(\to K\pi)\mu^+\mu^-$~\cite{Aaij:2013iag,Aaij:2016flj} process. In particular, the LHCb measurements of the ratios $R_K^{\rm exp}=0.745_{-0.074}^{+0.090}\pm0.036$ for $1.0\GeV^2\leq q^2\leq 6.0\GeV^2$~\cite{Aaij:2014ora} and $R_{K^*}^{\rm exp}=0.69_{-0.07}^{+0.11}\pm 0.05$ for $1.1\GeV^2\leq q^2\leq 6.0\GeV^2$~\cite{Aaij:2017vbb}, 
deviate from the SM predictions $R_{K^{(*)}}^{\rm SM}\simeq1$~\cite{Hiller:2003js,Bordone:2016gaq}~$2.6\sigma$ and $2.5\sigma$  respectively. 
These discrepancies in the ratios  $R_{K^{(*)}}$ indicate a violation of lepton flavour universality (LFU), if confirmed, which would be a clear signal  of NP~\cite{Li:2018lxi,Bifani:2018zmi}.

The baryonic decay $\Lambda_b \to \Lambda(\to N\pi) \ell^+ \ell^-$ \cite{Boer:2014kda}, which are mediated by the quark-level $b\to s\ell^+\ell^-$ transitions, has the potential to shed new light on the above-mentioned anomalies. Both $B\to K^{*}(\to K\pi)\ell^+\ell^-$ ~\cite{Altmannshofer:2008dz,Gratrex:2015hna} and $\Lambda_b \to \Lambda(\to N\pi) \ell^+ \ell^-$ decays can provide a number of angular observables, while the theoretical description of $\Lambda_b \to \Lambda(\to N\pi) \ell^+ \ell^-$ is cleaner than that of $B\to K^{*}(\to K\pi)\ell^+\ell^-$, because the $\Lambda$ is stable under the strong interactions. In addition, the subsequent weak decay $\Lambda\to N\pi$ in $\Lambda_b \to \Lambda \ell^+ \ell^-$ is parity violating, while the strong decay $K^*\to K\pi$ in $B\to K^*\ell^+\ell^-$ is not. Theoretical challenges in the process of $\Lambda_b \to \Lambda \ell^+ \ell^-$ are the evaluation of the hadronic $\Lambda_b \to \Lambda$ transition form factors. The form factors have been calculated by many methods, such as quark models~\cite{Cheng:1995fe,Mott:2011cx}, perturbative QCD (pQCD)~\cite{He:2006ud},  soft-collinear effective theory (SCET)~\cite{Wang:2011uv,Feldmann:2011xf}, light cone sum rules (LCSR)~\cite{Wang:2008sm,Wang:2015ndk} and lattice QCD~\cite{Detmold:2012vy,Detmold:2016pkz}. For the form factors involved, we use the  latest results from the calculation of lattice QCD \cite{Detmold:2016pkz}, which are extrapolated to the whole $q^2$ region using the Bourrely-Caprini-Lellouch (BCL) parametrization with $2+1$ flavor dynamics.  We refer to Ref.~\cite{Detmold:2016pkz} for recent reviews.

In the present theoretical study, the differential decay rate and the lepton-side forward-backward asymmetry in $\Lambda_b\to\Lambda\ell^+\ell^-$ have been studied for an incomplete list \cite{Aliev:2010uy,Chen:2001zc,Chen:2002rg,Aliev:2002ww,Azizi:2013eta,Wang:2016dne,Faustov:2017wbh}. 
The $\Lambda_b \to \Lambda(\to N\pi) \ell^+ \ell^-$ decay has been investigated theoretically in Refs.~\cite{Gutsche:2013pp,Boer:2014kda,Nasrullah:2018puc,Nasrullah:2018vky,Hu:2017qxj,Roy:2017dum,Blake:2017une,Das:2018iap}  for the angular observables. These current theoretical calculations do not consider tensor operators and lepton mass simultaneously. 
Experimentally, the $\Lambda_b\to\Lambda \mu^+\mu^-$ decay was measured firstly by CDF collaboration  with 24 signal events and statistical significance of 5.8 Gaussian standard deviations \cite{Aaltonen:2011qs}, corresponding to 6.8 ${\rm fb}^{-1}$ data. Soon afterward, the LHCb collaboration~\cite{Aaij:2013mna,Aaij:2015xza} has measured the differential branching , the lepton-side and  hadron-side forward-backward asymmetries. 
The crucial progresses in theory and experiment in recent years bring about bright prospects for the research of $\Lambda_b \to \Lambda(\to N\pi) \ell^+ \ell^-$ decay.

Given all of these motivations, we have recalculated the $\Lambda_b \to \Lambda(\to N\pi) \ell^+ \ell^-$ decay,  including the tensor operators for the first time, with unpolarized $\Lambda_b$ baryons and  massive leptons in this work. Since the $\Lambda_b$ polarization has been measured to be small by LHCb collaboration \cite{Aaij:2013oxa}, and the polarization effect will be averaged out in ATLAS and CMS detectors, we adopt unpolarized $\Lambda_b$ baryons in our calculations \cite{Hu:2017qxj}. Additionally, it is well known that branching fractions are susceptible to the uncertainties arising owing to the form factors and CKM matrix. Compared to the branching fractions, the ratios $R$ has the advantages as follows: the experimental systematic uncertainties are significantly reduced, the CKM matrix elements $V_{tb}V_{ts}$  cancel out and the sensitivity to $\Lambda_b\to \Lambda$ transition form factors becomes much weaker \cite{Yan:2019hpm}. For more accurate numerical results of LFU  ratios $R$, massive leptons are taken into account in our calculations.
All calculations are in Dirac representation in this work.
 
The article is organised as follows. In next section, we will give a brief overview of the effective Hamiltonian for $b\to s \ell^+\ell^-$ transitions. From this, we compute the hadronic helicity amplitudes in Sec.~\ref{sec:HLamb2Lam}. Then, we present the angular distribution of the four body  $\Lambda_b \to \Lambda(\to N \pi) \ell^+\ell^-$ decay in Sec.~\ref{sec:4bodyformula}. The phenomenological discussion is given in Sec.~\ref{sec:numLamb2Lam}, before drawing brief conclusions in Sec.~\ref{sec:conLamb2Lam}. Computational conventions and leptonic helicity amplitudes are collected in appendices.
\section{Effective Hamiltonian for $b\to s\ell^+ \ell^-$ transitions}
\label{sec:eff}
The $b \to s \ell^+ \ell^-$~transitions can be described by effective Hamiltonian which consists of two distinct parts, the short distance contributions contained in the Wilson coefficients $\wilson{i}(\mu)$ and the long distance physics described by the operator matrix elements $\op{i}$ . The NP model-independent effective Hamiltonian responsible for $b \to s \ell^+ \ell^-$~transitions can be written as~\cite{Buchalla:1995vs,Chetyrkin:1996vx,Buras:1998raa,Altmannshofer:2008dz,Boer:2014kda}:
\begin{equation}
    {\cal H}_{eff}
    = \frac{4 G_{ F}}{\sqrt 2} \, V_{tb} V_{ts}^* \,
    \frac{\alpha_e}{4 \pi} \, \sum_{i}\wilson{i}(\mu) \, \op{i}
    \,~,
\end{equation}
with the operators of dimension six 
\begin{align}
    \op{\rm S(')}
    & = [\bar s P_{\rm R(L)} b][\bar\ell \ell]\,, &
    \op{\rm P(')}
    & = [\bar s P_{\rm R(L)} b][\bar\ell \gamma_5\ell]\,,\nn\\
    \op{9(')}
    & = [\bar s \gamma^\mu P_{\rm L(R)} b][\bar\ell \gamma_\mu \ell]\,, &
    \op{10(')}
    & = [\bar s \gamma^\mu P_{\rm L(R)} b][\bar\ell \gamma_\mu \gamma_5 \ell]\,,\nn\\
    \op{\rm T}
    & = [\bar s \sigma^{\mu\nu} b] [\bar\ell \sigma_{\mu\nu}\ell]\,, &
    \op{\rm T5}
    & = -\frac{i}{2}  \epsilon_{\mu\nu\alpha\beta} \,
    [\bar s \sigma^{\mu\nu} b] [\bar\ell \sigma^{\alpha\beta}\ell]\,, \nn\\
    \op{7(')}
    &= \frac{m_{b}}{e} [\bar s \sigma^{\mu\nu} P_{\rm R(L)} b ] F_{\mu\nu}\,,
\end{align}
where $P_{\rm L,R}=(1\mp \gamma_{5})/2$, $\sigma_{\mu\nu}=i/2(\gamma_\mu\gamma_\nu-\gamma_\nu\gamma_\mu)$\footnote{The convention $\sigma_{\mu \nu}\gamma_{5} =-(i/2)\epsilon^{\mu\nu\alpha\beta}\sigma_{\alpha\beta}$ is used in this work, where $\epsilon^{0123}=-1$. } and $m_b$ denotes the running $b$ quark mass in the $\overline{\text{MS}}$ scheme. In the SM, the $b \to s \ell^+ \ell^-$ process contains the contribution of the electromagnetic operator $\op{7}$ and the semi-leptonic operators $\op{9,10}$. The new physical effects are achieved by changing the corresponding Wilson coefficients. 

 All nonlocal hadronic matrix elements are absorbed via the ``~effective Wilson coefficients~" $\wilson[\rm eff]{7}(q^2)$ and $\wilson[\rm eff]{9}(q^2)$. Following Refs.~\cite{Du:2015tda,Detmold:2016pkz}, we set the effective Wilson coefficients to
\begin{align}
C_7^{\rm eff}(q^2) = &  C_7 - \frac{1}{3} \big( C_3 + \frac{4}{3}\,C_4 + 20\,C_5  + \frac{80}{3}\, C_6 \big)\cr
&-\frac{\alpha_s}{4 \pi} \big[ \left(C_1 - 6\,C_2\right) F_{1,c}^{(7)}(q^2) + C_8\, F_8^{(7)}(q^2) \big]
\,,\cr
C_9^{\rm eff}(q^2) = &  C_9 + \frac{4}{3}\, C_3 + \frac{64}{9}\, C_5 + \frac{64}{27}\, C_6
+ h(0,q^2) \big( -\frac{1}{2}\, C_3 - \frac{2}{3}\, C_4 - 8\, C_5 - \frac{32}{3}\, C_6 \big) \nn\\
& + h(m_b,q^2) \big( -\frac{7}{2}\, C_3 - \frac{2}{3}\, C_4 - 38\, C_5 - \frac{32}{3}\, C_6 \big)\cr  
&+ h(m_c,q^2) \big( \frac{4}{3}\, C_1 + C_2 + 6\, C_3 + 60\, C_5 \big)  \cr 
& - \frac{\alpha_s}{4 \pi} \big[ C_1\, F_{1,c}^{(9)}(q^2) + C_2\, F_{2,c}^{(9)}(q^2) + C_8\, F_8^{(9)}(q^2) \big]\,, \nn
\end{align}
where the function $h(m_q,q^2)$ is~\cite{Beneke:2001at}:
\begin{align}
h(m_q, q^2) =&\frac{8}{27} -\frac{8}{9}\ln\frac{m_q}{m_b}  + \frac{4}{9}x - \frac{2}{9}(2+x)|1-x|^{1/2} \left\{
\begin{array}{ll}
\ln\left| \frac{\sqrt{1-x} + 1}{\sqrt{1-x} - 1}\right| - i\pi\,, & x \equiv \frac{4 m_c^2}{ q^2} < 1\,,  \\ & \\ 2 \arctan \frac{1}{\sqrt{x-1}}\,, & x \equiv \frac {4 m_c^2}{ q^2} > 1 \,,
\end{array}\right. \nn \\
h(0,q^2) =& \frac{8}{27} - \frac{4}{9} \ln\frac{q^2}{m_b^2}  + \frac{4}{9} i\pi \,.\nn
\end{align}
The functions~$F_{8}^{(7,9)}(q^2)$~ are taken form Ref.~\cite{Beneke:2001at}, the functions $F_{1,c}^{(7,9)}(q^2)$ and $F_{2,c}^{(7,9)}(q^2)$ are defined in Ref.~\cite{Asatryan:2001zw} for low $q^2$ and Ref.~\cite{Greub:2008cy} for high $q^2$. We evaluate the effective Wilson coefficients using the Mathematica packages provided by Ref. \cite{Greub:2008cy}. The Wilson coefficients of the SM are shown in Tab. 6 of Ref. \cite{Detmold:2016pkz}, with NNLL accuracy at the energy scale $\mu_b=4.2\GeV$. The charm and bottom masses appearing in the functions $h(m_q,q^2)$, $F_{1,c}^{(7,9)}(q^2)$ and $F_{8}^{(7,9)}(q^2)$ are defined in the pole scheme ~\cite{Greub:2008cy,Asatryan:2001zw}. 

In order to calculate the interference between operators conveniently, it is necessary to recombine and classify operators,
\begin{align}
\sum\wilson{i}(\mu) \, \op{i}=&(\wilson{7}+\wilson[{\rm NP}]{7})\op{7}+\wilson{7'}\op{7'}+(\wilson{9}+\wilson[{\rm NP}]{9})\op{9}+\wilson{9'}\op{9'}+(\wilson{10}+\wilson[{\rm NP}]{10})\op{10}\nn\\
&+\wilson{10'}\op{10'}+\wilson{\rm S}\op{\rm S}+\wilson{\rm S'}\op{\rm S'}+\wilson{\rm P}\op{\rm P}+\wilson{\rm P'}\op{\rm P'}+\wilson{\rm T}\op{\rm T}+\wilson{\rm T5}\op{\rm T5}\nn\\
=&\frac{1}{2}\big[(\wilson{\rm S}+\wilson{\rm S'})\bar s b \bar\ell\ell+\frac{-2m_b}{q^2}(\wilson{7}+\wilson[{\rm NP}]{7}+\wilson{7'})\bar s \tred{i}\sigma^{\mu\nu}q_\nu b\bar\ell\gamma_\mu\ell\cr
&+(\wilson{\rm S}-\wilson{\rm S'})\bar s\gamma_5 b\bar\ell\ell+\frac{-2m_b}{q^2}(\wilson{7}+\wilson[{\rm NP}]{7}-\wilson{7'})\bar s \tred{i} \sigma^{\mu\nu}q_\nu\gamma_5 b\bar\ell\gamma_\mu\ell\nn\\
&+(\wilson{\rm P}+\wilson{\rm P'})\bar s b\bar\ell\gamma_5\ell+(\wilson{9}+\wilson[{\rm NP}]{9}+\wilson{9'})\bar s \gamma^\mu b\bar\ell\gamma_\mu\ell\cr
&+(\wilson{\rm P}-\wilson{\rm P'})\bar s\gamma_5 b\bar\ell\gamma_5\ell+(-\wilson{9}-\wilson[{\rm NP}]{9}+\wilson{9'})\bar s \gamma^\mu\gamma_5 b\bar\ell\gamma_\mu\ell\nn\\
&+2\wilson{\rm T5}\bar s \tblue{i}\sigma^{\mu\nu} b\bar\ell(\tblue{-i}) \sigma_{\mu\nu}\gamma_5 \ell+(-\wilson{10}-\wilson[{\rm NP}]{10}+\wilson{10'})\bar s \gamma^\mu\gamma_5 b\bar\ell\gamma_\mu\gamma_5 \ell\nn\\
&+2\wilson{\rm T}\bar s \tblue{i}\sigma^{\mu\nu} b\bar\ell(\tblue{-i}) \sigma_{\mu\nu}\ell+(\wilson{10}+\wilson[{\rm NP}]{10}+\wilson{10'})\bar s \gamma^\mu b\bar\ell\gamma_\mu\gamma_5 \ell\big]\, . 
\end{align}
Notice that the electromagnetic operators~$\op{7(')}$~contribute to the $b\to s\ell^+ \ell^-$ transitions through photon exchange
\begin{align} \label{eq:o7}
    \bra{\Lambda(k) \ell^+(q_1) \ell^-(q_2)} \op{7(')} \ket{\Lambda_b(p)}
    & = -\frac{2 m_b}{q^2} \, \bra{\Lambda} \bar{s} \, \tred{i}\sigma^{\mu\nu} q_\nu \, P_{\rm R(L)} b \ket{\Lambda_b} \,
    [\bar{u_\ell} \gamma_\mu v_\ell]\,,
\end{align}
where $q^\mu=p^\mu-k^\mu$, which is the momentum transfer to the lepton pair.

For simplicity of the hadronic helicity amplitudes, we redefine Wilson coefficients as follows 
\begin{align}
\wilson[\pm]{\rm S} =&\wilson{\rm S}\pm\wilson{\rm S'}\, ,&
\wilson[\pm]{\rm P} =&\wilson{\rm P}\pm\wilson{\rm P'}\, ,\nn\\
\wilson[\pm]{7}=&\frac{-2m_b}{q^2}(\wilson{7}+\wilson[{\rm NP}]{7}\pm\wilson{7'})\, ,&
\wilson[\pm]{9}=&\pm\wilson{9}\pm\wilson[{\rm NP}]{9}+\wilson{9'}\,  ,\nn\\
\wilson[\pm]{10}=&\pm\wilson{10}\pm\wilson[{\rm NP}]{10}+\wilson{10'}\, .&
\end{align}

\section{ Hadronic Matrix Elements}\label{sec:HLamb2Lam}
\subsection{ $\Lambda_b\to \Lambda$~hadronic helicity amplitudes}
The $\Lambda_b\rightarrow\Lambda$ hadronic matrix elements can be expressed as a series of  form factors~\cite{Detmold:2016pkz}. In order to calculate the helicity amplitudes, we have used explicit expressions for the Dirac spinors which are defined in App. A of Ref.~\cite{Yan:2019hpm}. The definitions of hadronic helicity amplitudes are collected in the App. \ref{sec:helicity}, which are consistent with Ref.~\cite{Yan:2019hpm}.  Using the above mentioned relationships, the non-zero helicity amplitudes for $\Lambda_b\to \Lambda$~decay can be written as:
\begin{align}\label{H:Lamb2Lam}
H_{\pm 1/2,\pm 1/2}^{\rm P}=& \mp \frac{m_{\Lambda_b}+m_{\Lambda}}{m_{b}+m_{s}}G_{0}\sqrt{Q_{-}}\, ,
&H_{\pm 1/2,\pm 1/2}^{\rm S}=&  \frac{m_{\Lambda_b}-m_{\Lambda}}{m_{b}-m_{s}}F_{0}\sqrt{Q_{+}}\, ,\cr 
H_{\pm 1/2,\pm 1/2,0}^{\rm V}=& \frac{\sqrt{Q_-}}{\sqrt{q^2}}(m_{\Lambda_b}+m_{\Lambda})F_{+} \, ,
&H_{\pm 1/2,\pm 1/2,t}^{\rm V}=& \frac{\sqrt{Q_+}}{\sqrt{q^2}}(m_{\Lambda_b}-m_{\Lambda})F_{0} \, , \cr 
H_{\pm 1/2,\pm 1/2,t}^{\rm A}=&\pm  \frac{\sqrt{Q_-}}{\sqrt{q^2}} (m_{\Lambda_b}+m_{\Lambda}) G_0\, ,
&H_{\mp 1/2,\pm 1/2,\pm }^{\rm V}=&  \sqrt{2Q_-} F_\perp\, ,\cr 
H_{\pm 1/2,\pm 1/2,0}^{\rm A}=&\pm  \frac{\sqrt{Q_+}}{\sqrt{q^2}} (m_{\Lambda_b}-m_{\Lambda}) G_+ \, ,
&H_{\mp 1/2,\pm 1/2,\pm }^{\rm A}=&\pm  \sqrt{2Q_+}G_\perp\, ,\cr 
H_{\mp 1/2,\pm 1/2,\pm }^{\rm Tq}=&- \sqrt{2}\sqrt{Q_-}(m_{\Lambda_b}+m_{\Lambda})h_\perp \, ,
&H_{\pm 1/2,\pm 1/2,0}^{\rm Tq}=&- \sqrt{q^2 Q_-}h_+ \, ,\cr 
H_{\mp 1/2,\pm 1/2,\pm }^{\rm Tq5}=&\pm  \sqrt{2}\sqrt{Q_+}(m_{\Lambda_b}-m_{\Lambda})\widetilde{h}_\perp \, ,
&H_{\pm 1/2,\pm 1/2,0}^{\rm Tq5}=&\pm  \sqrt{q^2 Q_+} \widetilde{h}_+\, , \cr 
H_{\mp 1/2,0,\mp}^{{\rm T},\pm 1/2}=& \frac{\sqrt{2Q_+}}{\sqrt{q^2}}\widetilde{h}_\perp (m_{\Lambda_b}-m_{\Lambda}) \, ,
&H_{\pm 1/2,t,0}^{{\rm T},\pm 1/2}=& h_+\sqrt{Q_-} \, , \cr 
H_{\mp 1/2,t,\mp}^{{\rm T},\pm 1/2}=& \frac{\sqrt{2Q_-}}{\sqrt{q^2}}h_\perp (m_{\Lambda_b}+m_{\Lambda}) \, ,
& H_{\pm 1/2,+,-}^{{\rm T},\pm 1/2}=&\mp \widetilde{h}_+\sqrt{Q_+}\,. 
\end{align}
The decay amplitude is analysed by performing a decomposition according to helicity amplitudes, which explicated in Ref.~\cite{Haber:1994pe}. The conceptual foundation of helicity amplitudes is the orthonormality and completeness relation~\cite{Korner:1989qb}
	\begin{align}\label{eq:orthonormality and completeness}
	\sum\limits_\mu \epsilon_{\mu}^{*}(m)\epsilon^{\mu}(n)=g_{mn}\, ,
	\qquad
	\sum\limits_{m,n} \epsilon_{\mu}(m)\epsilon_{\nu}^{*}(n)g_{mn}=g_{\mu\nu}\, ,
	\qquad
	m,n\in\{t,\pm,0\}\, ,
	\end{align}
	where $g_{mn}=\text{diag}(+1,-1,-1,-1)$.  
	As polarization vectors satisfying orthogonal normalization relation are not unique, the form of hadronic and leptonic helicity amplitudes may be various under different calculation conventions. 

\subsection{ $\Lambda\to N\pi$~decay amplitudes}
~$\Lambda  \to N\pi$~decay is governed by the $\Delta S=1$ effective Hamiltonian~\cite{Boer:2014kda}
\begin{align}
    H^\text{eff}_{\Delta S = 1} =N_2
    \left[\bar{d} \gamma_\mu P_L u\right]\left[\bar{u} \gamma^\mu P_L s\right] \,,
\end{align}
where $N_2={2\sqrt{2} G_F} \, V_{ud}^* V_{us}$. 
The hadronic matrix elements of $\Lambda  \to N\pi$~decay are parametrized in terms of two hadronic parameters $\xi$ and $\omega$ as~\cite{PDG:2018}
\begin{align}
&
\bra{p(k_1, s_N) \pi^-(k_2)} \left[\bar{d} \gamma_\mu P_L u\right]\left[\bar{u} \gamma^\mu P_L s\right]
\ket{\Lambda(k, s_\Lambda)} 
\cr & 
= \big[\bar u(k_1, s_N) \big(\tred{-}\xi \,\gamma_5 + \omega\big) u(k, s_\Lambda)\big] \equiv H_2(s_\Lambda, s_N)\,,
\end{align}
where $\xi$ and $\omega$ can be extracted from the $\Lambda\to p\pi^-$ decay width and polarization measurements, corresponding to A and B of Ref.~\cite{PDG:2018} respectively. 

According to the discussion in the App. A of Ref.~\cite{Yan:2019hpm}, we take $u(k_1, s_N),~u(k, s_\Lambda)$ as
\begin{align}
u(k_1, s_N)=\begin{spmatrix}
\hphantom{2s_N}\frac{\sqrt{r_+}}{\sqrt{m_{\Lambda}}}~\chi(\vec{k}_1,s_N)\\ 2s_N\frac{\sqrt{r_-}}{\sqrt{m_{\Lambda}}} ~\chi(\vec{k}_1,s_N)
\end{spmatrix},\quad~u(k, s_\Lambda)=\sqrt{2m_{\Lambda}}~\begin{spmatrix}
\chi(\vec{k},s_\Lambda)\\ 0
\end{spmatrix}\, ,
\end{align}
where
\begin{align}
\chi(\vec{k}_1,+\frac{1}{2})=&\begin{spmatrix}\hphantom{e^{i\phi}}\cos\frac{\theta_{\Lambda}}{2} \\ e^{i\phi}\sin\frac{\theta_{\Lambda}}{2} \end{spmatrix}\, ,&\chi(\vec{k}_1,-\frac{1}{2})=&\begin{spmatrix}-e^{-i\phi}\sin\frac{\theta_{\Lambda}}{2} \\ \hphantom{-e^{-i\phi}}\cos\frac{\theta_{\Lambda}}{2} \end{spmatrix}\, ,\cr
\chi(\vec{k},+\frac{1}{2})=&\begin{spmatrix}1 \\ 0 \end{spmatrix}\, ,&\chi(\vec{k},-\frac{1}{2})=&\begin{spmatrix}0 \\ 1 \end{spmatrix}\, .
\end{align}

Using the above spinor conventions, the following helicity amplitudes can be easily obtained, 
\begin{align}
    H_2(+1/2,+1/2) =& \left(\sqrt{r_+} \, \omega + \sqrt{r_-} \, \xi \right) \cos\frac{\theta_\Lambda}{2} \,, \cr
    H_2(+1/2,-1/2) =& \left(-\sqrt{r_+} \, \omega + \sqrt{r_-} \, \xi \right) \sin\frac{\theta_\Lambda}{2}\,
    e^{i \phi}  \,, \cr
    H_2(-1/2,+1/2) =& \left(\sqrt{r_+} \, \omega + \sqrt{r_-} \, \xi \right) \sin\frac{\theta_\Lambda}{2}\,
    e^{-i \phi} \,, \cr
    H_2(-1/2,-1/2) =& \left(\sqrt{r_+}\, \omega - \sqrt{r_-} \,\xi \right) \cos\frac{\theta_\Lambda}{2} \,,
\end{align}
with 
\begin{align}
    r_{\pm} \equiv (m_{\Lambda} \pm m_N)^2 - m_{\pi}^2 \,.
\end{align}


The~$\Lambda  \to N\pi$~decay width can be defined as
\begin{align}
    \Gamma_2(s_{\Lambda}^{(a)},s_{\Lambda}^{(b)}) =  |N_2|^2 \frac{\sqrt{r_+ r_-}}{16 \pi m_{\Lambda}^3} \,
    \sum_{s_N} H_2(s_{\Lambda}^{(a)},s_N) \, H_2^* (s_{\Lambda}^{(b)},s_N) \,,
\end{align}
by which it can be obtained that
\begin{align}
     \Gamma_2(+1/2, +1/2) =&  (1 + \alpha  \, \cos\theta_\Lambda) \, \Gamma_\Lambda\,,
     &\Gamma_2(+1/2, -1/2) =& \alpha \, \sin\theta_\Lambda \, e^{i \phi} \, \Gamma_\Lambda \,,\cr
     \Gamma_2(-1/2, -1/2) =& (1 - \alpha \, \cos\theta_\Lambda) \Gamma_\Lambda\,  ,
     &\Gamma_2(-1/2, +1/2) =& \alpha \, \sin\theta_\Lambda e^{-i \phi}\,\Gamma_\Lambda \,.
\end{align}
The total decay width of~$\Lambda \to N\pi$~is given by
\begin{align}
    \Gamma_\Lambda  =& \frac{1}{2}[\Gamma_2(+1/2, +1/2) + \Gamma_2(-1/2, -1/2)]\cr
                    =& \frac{|N_2|^2 \sqrt{r_+ r_-}}{16 \pi m_\Lambda^3} \left(
                    r_- \,|\xi|^2 + r_+ \, |\omega|^2 \right)\,,
\end{align}
and the parity-violating decay parameter $\alpha$~reads~\cite{PDG:2018}
\begin{align}
    \alpha = \frac{2 \,\real[\omega^* \,\xi]}{\sqrt{\frac{r_-}{r_+}} \, |\xi|^2 + \sqrt{\frac{r_+}{r_-}} \, |\omega|^2}\, .
\end{align}

\section{ Angular Observables}\label{sec:4bodyformula}
\subsection{Four body angular distribution}
According to the conventions of Ref.~\cite{Boer:2014kda}, we can define the four body angular distribution as a four body differential decay width 
\begin{align}\label{eq:defK}
  \overline{K}(q^2, \cos\theta_\ell, \cos\theta_\Lambda, \phi) \equiv
  \frac{8\pi}{3} \frac{\dd^4 \Gamma}{\dd q^2\,\dd \cos\theta_\ell\,\dd \cos\theta_\Lambda\,\dd \phi} \,~,
\end{align}
where $\overline{K}(q^2, \cos\theta_\ell, \cos\theta_\Lambda, \phi)$  can be decomposed in terms of a set of trigonometric functions and angular coefficients 
\begin{align}
    \label{eq:angular-distribution}
    \overline{K}(q^2, \cos\theta_\ell, \cos\theta_\Lambda, \phi)
    & =
         \big( \overline{K}_{1ss} \sin^2\theta_\ell +\, \overline{K}_{1cc} \cos^2\theta_\ell + \overline{K}_{1c} \cos\theta_\ell\big) \,\cr
    &  + \big( \overline{K}_{2ss} \sin^2\theta_\ell +\, \overline{K}_{2cc} \cos^2\theta_\ell + \overline{K}_{2c} \cos\theta_\ell\big) \cos\theta_\Lambda
    \cr
    &  + \big( \overline{K}_{3sc}\sin\theta_\ell \cos\theta_\ell + \overline{K}_{3s} \sin\theta_\ell\big) \sin\theta_\Lambda \sin\phi\cr
    &  + \big( \overline{K}_{4sc}\sin\theta_\ell \cos\theta_\ell + \overline{K}_{4s} \sin\theta_\ell\big) \sin\theta_\Lambda \cos\phi \,~ .
\end{align}
For the convenience, we have the following conventions, 
\begin{align}
N\equiv & G_F V_{tb}V_{ts}^{*}\alpha_e\sqrt{\frac{q^2\sqrt{\lambda(m_{\Lambda_b}^2,m_{\Lambda}^2,q^2)}}{3\times 2^{11}m_{\Lambda_b}^3 \pi^5 }(1-\frac{4m_{\ell}^2}{q^2})}\, ~,\cr
\overline{K}_{\{\cdots\}}\equiv & 2\abs{N}^2 K_{\{\cdots\}} \,~ ,\cr
K_{\{\cdots\}}=&K_{\{\cdots\}}^{\rm SP}+K_{\{\cdots\}}^{\rm VA}+K_{\{\cdots\}}^{\rm Tq}+K_{\{\cdots\}}^{\rm T}+K_{\{\cdots\}}^{\rm SP-VA}+K_{\{\cdots\}}^{\rm SP-T}\cr
&+K_{\{\cdots\}}^{\rm SP-T}+K_{\{\cdots\}}^{\rm VA-Tq}+K_{\{\cdots\}}^{\rm VA-T}+K_{\{\cdots\}}^{\rm Tq-T}\,~ .
\end{align}
Utilizing the formulas in App.~\ref{sec:4body}, we can get the explicit expressions of $K_{\{\cdots\}}$\,. 
The nonzero angular coefficients of the scalar and pseudo-scalar operators read 
\begin{align}
K_{1cc}^{\rm SP} =&(\abs{\wilson[-]{\rm P}}^2+\beta_{\ell}^2 \abs{\wilson[-]{\rm S}}^2)(H_{1/2,1/2}^{\rm P})^2+(\abs{\wilson[+]{\rm P}}^2+\beta_{\ell}^2 \abs{\wilson[+]{\rm S}}^2)(H_{1/2,1/2}^{\rm S})^2  \,,\cr
K_{1ss}^{\rm SP} =& K_{1cc}^{\rm SP}\,,\cr
K_{2cc}^{\rm SP} =& 2\alpha\big[\real[\wilson[+]{\rm P}\wilson[-*]{\rm P}]+\beta_{\ell}^2 \real [\wilson[+]{\rm S}\wilson[-*]{\rm S}]\big] H_{1/2,1/2}^{\rm P}H_{1/2,1/2}^{\rm S}  \,,\cr
K_{2ss}^{\rm SP} =& K_{2cc}^{\rm SP}  \,.
\end{align}
The nonzero angular coefficients of the vector and axial-vector operators are 
\begin{align}
K_{1c}^{\rm VA}  =&  -2\beta_{\ell}\real[\wilson[+]{9}\wilson[-*]{10}+\wilson[-]{9}\wilson[+*]{10}] H_{-1/2,1/2,+}^{\rm A}H_{-1/2,1/2,+}^{\rm V} \, ,\cr
K_{1cc}^{\rm VA} =&  (\abs{\wilson[-]{9}}^2+\beta_{\ell}^2\abs{\wilson[-]{10}}^2 \big](H_{-1/2,1/2,+}^{\rm A})^2 +\big[\abs{\wilson[+]{9}}^2+\beta_{\ell}^2\abs{\wilson[+]{10}}^2 \big](H_{-1/2,1/2,+}^{\rm V})^2 \cr
&+\frac{4m_{\ell}^2}{q^2}\big[\abs{\wilson[-]{9}}^2(H_{1/2,1/2,0}^{\rm A})^2+\abs{\wilson[-]{10}}^2(H_{1/2,1/2,t}^{\rm A})^2 \cr
&+\abs{\wilson[+]{9}}^2(H_{1/2,1/2,0}^{\rm V})^2+\abs{\wilson[+]{10}}^2(H_{1/2,1/2,t}^{\rm V})^2)  \, ,\cr
K_{1ss}^{\rm VA} =& (\beta_{\ell}^2\abs{\wilson[-]{10}}^2+\abs{\wilson[-]{9}}^2 )(H_{1/2,1/2,0}^{\rm A})^2+\frac{1}{2}(\beta_{\ell}^{2}\abs{\wilson[-]{10}}^2+\beta_{\ell+}^{2}\abs{\wilson[-]{9}}^2)(H_{-1/2,1/2,+}^{\rm A})^2\cr
&+(\beta_{\ell}^{2}\abs{\wilson[+]{10}}^2+\abs{\wilson[+]{9}}^2)(H_{1/2,1/2,0}^{\rm V})^2+\frac{1}{2}(\beta_{\ell}^2\abs{\wilson[+]{10}}^2+\beta_{\ell+}^{2}\abs{\wilson[+]{9}}^2 )(H_{-1/2,1/2,+}^{\rm V})^2 \cr
&+\frac{4m_{\ell}^2}{q^2}\big[\abs{\wilson[-]{10}}^2(H_{1/2,1/2,t}^{\rm A})^2+\abs{\wilson[+]{10}}^2(H_{1/2,1/2,t}^{\rm V})^2\big] \, , \cr
K_{2c}^{\rm VA}  =&  -2\alpha\beta_{\ell}\big[\real[\wilson[-]{9}\wilson[-*]{10}](H_{-1/2,1/2,+}^{\rm A})^2+\real[\wilson[+]{9}\wilson[+*]{10}](H_{-1/2,1/2,+}^{\rm V})^2\big] \,  ,\cr
K_{2cc}^{\rm VA} =& \frac{8\alpha m_{\ell}^2}{q^2}\big[\real[\wilson[+]{10}\wilson[-*]{10}]H_{1/2,1/2,t}^{\rm A}H_{1/2,1/2,t}^{\rm V}+\real[\wilson[+]{9}\wilson[-*]{9}]H_{1/2,1/2,0}^{\rm A}H_{1/2,1/2,0}^{\rm V} \big]  \cr
&+2\alpha \big[\beta_{\ell}^2\real[\wilson[+]{10}\wilson[-*]{10}]+\real[\wilson[+]{9}\wilson[-*]{9}]\big]H_{-1/2,1/2,+}^{\rm A} H_{-1/2,1/2,+}^{\rm V}\, ,\cr
K_{2ss}^{\rm VA} =& \alpha\big[\beta_{\ell+}^{2}\real[\wilson[+]{9}\wilson[-*]{9}]+\beta_{\ell}^2\real[\wilson[+]{10}\wilson[-*]{10}]\big]H_{-1/2,1/2,+}^{\rm A}H_{-1/2,1/2,+}^{\rm V} \cr
&+2\alpha(\real[\wilson[+]{9}\wilson[-*]{9}]+\beta_{\ell}^2\real[\wilson[+]{10}\wilson[-*]{10}])H_{1/2,1/2,0}^{\rm A}H_{1/2,1/2,0}^{\rm V} \cr
&+\frac{8\alpha m_{\ell}^2}{q^2}\real[\wilson[+]{10}\wilson[-*]{10}]H_{1/2,1/2,t}^{\rm A}H_{1/2,1/2,t}^{\rm V} \, ,\cr
K_{3s}^{\rm VA}  =&  \sqrt{2}\alpha\beta_{\ell}~\im[\wilson[+]{9}\wilson[-*]{10}+\wilson[+]{10}\wilson[-*]{9}]\big[H_{1/2,1/2,0}^{\rm A}H_{-1/2,1/2,+}^{\rm V}+H_{1/2,1/2,0}^{\rm V}H_{-1/2,1/2,+}^{\rm A} \big]  \, ,\cr
K_{4sc}^{\rm VA} =&  \sqrt{2}\alpha\beta_{\ell}^2~\real[\wilson[+]{9}\wilson[-*]{9}+\wilson[+]{10}\wilson[-*]{10}]\big[H_{1/2,1/2,0}^{\rm A}H_{-1/2,1/2,+}^{\rm V}-H_{1/2,1/2,0}^{\rm V}H_{-1/2,1/2,+}^{\rm A} \big]  \, ,\cr
K_{4s}^{\rm VA}  =&  2\sqrt{2}\alpha\beta_{\ell}~\big[\real[\wilson[+]{9}\wilson[+*]{10}]H_{1/2,1/2,0}^{\rm V}H_{-1/2,1/2,+}^{\rm V} \cr
&-\real[\wilson[-]{9}\wilson[-*]{10}]H_{1/2,1/2,0}^{\rm A}H_{-1/2,1/2,+}^{\rm A}\big]  \, .
\end{align}
The nonzero angular coefficients of the electromagnetic operators~$\op{7(')}$  read as
\begin{align}
K_{1cc}^{\rm Tq} =&  \abs{\wilson[+]{7}}^2\big[(H_{-1/2,1/2,+}^{\rm Tq})^2+\frac{4  m_{\ell}^2}{q^2}(H_{1/2,1/2,0}^{\rm Tq})^2 \big]\cr
&+\abs{\wilson[-]{7}}^2\big[(H_{-1/2,1/2,+}^{\rm Tq5})^2+\frac{4  m_{\ell}^2}{q^2}(H_{1/2,1/2,0}^{\rm Tq5})^2 \big]   \, ,\cr
K_{1ss}^{\rm Tq} =&  \abs{\wilson[+]{7}}^2\big[(H_{1/2,1/2,0}^{\rm Tq})^2+\frac{\beta_{\ell+}^{2}}{2}(H_{-1/2,1/2,+}^{\rm Tq})^2 \big]\cr
&+\abs{\wilson[-]{7}}^2\big[(H_{1/2,1/2,0}^{\rm Tq5})^2+\frac{\beta_{\ell+}^{2}}{2}(H_{-1/2,1/2,+}^{\rm Tq5})^2\big]   \, ,\cr
K_{2cc}^{\rm Tq} =&  2\alpha \real[\wilson[+]{7}\wilson[-*]{7}]\big[H_{-1/2,1/2,+}^{\rm Tq}H_{-1/2,1/2,+}^{\rm Tq5}+\frac{4  m_{\ell}^2}{q^2}H_{1/2,1/2,0}^{\rm Tq}H_{1/2,1/2,0}^{\rm Tq5}\big]  \, ,\cr
K_{2ss}^{\rm Tq} =&  \alpha \real[\wilson[+]{7}\wilson[-*]{7}]\big[2H_{1/2,1/2,0}^{\rm Tq}H_{1/2,1/2,0}^{\rm Tq5}+\beta_{\ell+}^{2}H_{-1/2,1/2,+}^{\rm Tq}H_{-1/2,1/2,+}^{\rm Tq5}\big] \,  ,\cr
K_{4sc}^{\rm Tq} =&  \sqrt{2}\alpha\beta_{\ell}^2\real[\wilson[+]{7}\wilson[-*]{7}]\big[H_{1/2,1/2,0}^{\rm Tq5}H_{-1/2,1/2,+}^{\rm Tq}-H_{1/2,1/2,0}^{\rm Tq}H_{-1/2,1/2,+}^{\rm Tq5} \big] \,  .
\end{align}
The nonzero angular coefficients of the tensor operators $\op{\rm T,~T5}$ are as follows
\begin{align}
K_{1cc}^{\rm T} =&  16(\beta_{\ell}^2\abs{\wilson{\rm T}}^2+\abs{\wilson{\rm T5}}^2)(H_{1/2,+,-}^{{\rm T},1/2})^2 +16(\abs{\wilson{\rm T}}^2+\beta_{\ell}^2\abs{\wilson{\rm T5}}^2)(H_{1/2,t,0}^{{\rm T},1/2})^2\cr
&+\frac{64  m_{\ell}^2}{q^2}\big[\abs{\wilson{\rm T5}}^2(H_{-1/2,0,-}^{{\rm T},1/2})^2+\abs{\wilson{\rm T}}^2(H_{-1/2,t,-}^{{\rm T},1/2})^2\big]  \, ,\cr
K_{1ss}^{\rm T} =&  8(\beta_{\ell}^{2}\abs{\wilson{\rm T5}}^2+\beta_{\ell+}^{2}\abs{\wilson{\rm T}}^2)(H_{-1/2,t,-}^{{\rm T},1/2})^2 +8(\beta_{\ell+}^{2}\abs{\wilson{\rm T5}}^2+\beta_{\ell}^2\abs{\wilson{\rm T}}^2)(H_{-1/2,0,-}^{{\rm T},1/2})^2 \cr
&+\frac{64  m_{\ell}^2}{q^2}\big[\abs{\wilson{\rm T5}}^2(H_{1/2,+,-}^{{\rm T},1/2})^2+\abs{\wilson{\rm T}}^2(H_{1/2,t,0}^{{\rm T},1/2})^2\big]  \, ,\cr
K_{2cc}^{\rm T} =&  64\alpha \real[\wilson{\rm T5}\wilson[*]{\rm T}]\big[(1-\frac{2m_{\ell}^{2}}{q^2})H_{1/2,+,-}^{{\rm T},1/2}H_{1/2,t,0}^{{\rm T},1/2}-\frac{2m_{\ell}^{2}}{q^2}H_{-1/2,0,-}^{{\rm T},1/2}H_{-1/2,t,-}^{{\rm T},1/2} \big]   \, ,\cr
K_{2ss}^{\rm T} =&  32\alpha \real[\wilson{\rm T5}\wilson[*]{\rm T}]\big[\frac{4m_{\ell}^{2}}{q^2}H_{1/2,+,-}^{{\rm T},1/2}H_{1/2,t,0}^{{\rm T},1/2}-H_{-1/2,0,-}^{{\rm T},1/2}H_{-1/2,t,-}^{{\rm T},1/2} \big]  \, ,\cr
K_{4sc}^{\rm T} =&  -32\sqrt{2}\alpha\beta_{\ell}^2 \real[\wilson{\rm T5}\wilson[*]{\rm T}]\big[H_{-1/2,t,-}^{{\rm T},1/2}H_{1/2,+,-}^{{\rm T},1/2}+H_{-1/2,0,-}^{{\rm T},1/2}H_{1/2,t,0}^{{\rm T},1/2} \big]  \,.
\end{align}
The nonzero angular coefficients of  interference between the scale (pseudo-scalar) and vector (axial-vector) operators  can be  enumerated as follows
\begin{align}
K_{1c}^{\rm SP-VA}  =&  \frac{4\beta_{\ell} m_{\ell}}{\sqrt{q^2}}\big[\real[\wilson[-]{\rm S}\wilson[-*]{9}]H_{1/2,1/2}^{\rm P}H_{1/2,1/2,0}^{\rm A}+\real[\wilson[+]{\rm S}\wilson[+*]{9}]H_{1/2,1/2}^{\rm S}H_{1/2,1/2,0}^{\rm V}\big]   \,,\cr
K_{1cc}^{\rm SP-VA} =&  \frac{4 m_{\ell}}{\sqrt{q^2}}\big[\real[\wilson[-]{\rm P}\wilson[-*]{10}]H_{1/2,1/2}^{\rm P}H_{1/2,1/2,t}^{\rm A}+\real[\wilson[+]{\rm P}\wilson[+*]{10}]H_{1/2,1/2}^{\rm S}H_{1/2,1/2,t}^{\rm V}\big]  \, ,\cr
K_{1ss}^{\rm SP-VA} =&  K_{1cc}^{\rm SP-VA}\, ,\cr
K_{2c}^{\rm SP-VA}  =&  \frac{4\alpha\beta_{\ell} m_{\ell}}{\sqrt{q^2}}\big[\real[\wilson[+]{\rm S}\wilson[-*]{9}]H_{1/2,1/2}^{\rm S}H_{1/2,1/2,0}^{\rm A}+\real[\wilson[-]{\rm S}\wilson[+*]{9}]H_{1/2,1/2}^{\rm P}H_{1/2,1/2,0}^{\rm V}\big]   \,,\cr
K_{2cc}^{\rm SP-VA} =&  \frac{4\alpha m_{\ell}}{\sqrt{q^2}}\big[\real[\wilson[+]{\rm P}\wilson[-*]{10}]H_{1/2,1/2}^{\rm S}H_{1/2,1/2,t}^{\rm A}+\real[\wilson[-]{\rm P}\wilson[+*]{10}]H_{1/2,1/2}^{\rm P}H_{1/2,1/2,t}^{\rm V}\big]  \, ,\cr
K_{2ss}^{\rm SP-VA} =&  K_{2cc}^{\rm SP-VA} \, ,\cr
K_{3s}^{\rm SP-VA}  =&  \frac{2\sqrt{2}\alpha\beta_{\ell} m_{\ell}}{\sqrt{q^2}}\big[\im[\wilson[+]{\rm S}\wilson[+*]{9}]H_{1/2,1/2}^{\rm S}H_{-1/2,1/2,+}^{\rm V}-\im[\wilson[-]{\rm S}\wilson[-*]{9}]H_{1/2,1/2}^{\rm P}H_{-1/2,1/2,+}^{\rm A}\big]   \, ,\cr
K_{4s}^{\rm SP-VA}  =&  \frac{2\sqrt{2}\alpha\beta_{\ell} m_{\ell}}{\sqrt{q^2}}\big[\real[\wilson[+]{\rm S}\wilson[-*]{9}]H_{1/2,1/2}^{\rm S}H_{-1/2,1/2,+}^{\rm A}\cr
&-\real[\wilson[-]{\rm S}\wilson[+*]{9}]H_{1/2,1/2}^{\rm P}H_{-1/2,1/2,+}^{\rm V}\big]  \,  .
\end{align}
The nonzero angular coefficients of  interference between  the scale (pseudo-scalar) and electromagnetic operators can be written as
\begin{align}
K_{1c}^{\rm SP-Tq}  =&  \frac{4\beta_{\ell} m_{\ell}}{\sqrt{q^2}}\big[\real[\wilson[-]{\rm S}\wilson[-*]{7}]H_{1/2,1/2}^{\rm P}H_{1/2,1/2,0}^{\rm Tq5}+\real[\wilson[+]{\rm S}\wilson[+*]{7}]H_{1/2,1/2}^{\rm S}H_{1/2,1/2,0}^{\rm Tq} \big]  \, ,\cr
K_{2c}^{\rm SP-Tq}  =&  \frac{4\alpha\beta_{\ell} m_{\ell}}{\sqrt{q^2}}\big[\real[\wilson[+]{\rm S}\wilson[-*]{7}]H_{1/2,1/2}^{\rm S}H_{1/2,1/2,0}^{\rm Tq5}+\real[\wilson[-]{\rm S}\wilson[+*]{7}]H_{1/2,1/2}^{\rm P}H_{1/2,1/2,0}^{\rm Tq} \big]  \, ,\cr
K_{3s}^{\rm SP-Tq}  =&  \frac{2\sqrt{2}\alpha\beta_{\ell} m_{\ell}}{\sqrt{q^2}}\big[\im[\wilson[+]{\rm S}\wilson[+*]{7}]H_{1/2,1/2}^{\rm S}H_{-1/2,1/2,+}^{\rm Tq}-\im[\wilson[-]{\rm S}\wilson[-*]{7}]H_{1/2,1/2}^{\rm P}H_{-1/2,1/2,+}^{\rm Tq5} \big]  \, ,\cr
K_{4s}^{\rm SP-Tq}  =&  \frac{2\sqrt{2}\alpha\beta_{\ell} m_{\ell}}{\sqrt{q^2}}\big[\real[\wilson[+]{\rm S}\wilson[-*]{7}]H_{1/2,1/2}^{\rm S}H_{-1/2,1/2,+}^{\rm Tq5}\cr
&-\real[\wilson[-]{\rm S}\wilson[+*]{7}]H_{1/2,1/2}^{\rm P}H_{-1/2,1/2,+}^{\rm Tq} \big]  \, .
\end{align}
The nonzero angular coefficients of  interference between  the scale (pseudo-scalar) and tensor operators can be listed as
\begin{align}
K_{1c}^{\rm SP-T}  =&  8\beta_{\ell}\big[\real[\wilson{\rm T}\wilson[-*]{\rm P}+\wilson{\rm T5}\wilson[-*]{\rm S}]H_{1/2,1/2}^{\rm P}H_{1/2,+,-}^{{\rm T},1/2}\cr
&+\real[\wilson{\rm T}\wilson[+*]{\rm S}+\wilson{\rm T5}\wilson[+*]{\rm P}]H_{1/2,1/2}^{\rm S}H_{1/2,t,0}^{{\rm T},1/2} \big] \, , \cr
K_{2c}^{\rm SP-T}  =&  8\alpha\beta_{\ell}\big[\real[\wilson{\rm T}\wilson[-*]{\rm S}+\wilson{\rm T5}\wilson[-*]{\rm P}]H_{1/2,1/2}^{\rm P}H_{1/2,t,0}^{{\rm T},1/2}\cr
&+\real[\wilson{\rm T}\wilson[+*]{\rm P}+\wilson{\rm T5}\wilson[+*]{\rm S}]H_{1/2,1/2}^{\rm S}H_{1/2,+,-}^{{\rm T},1/2} \big]  \, ,\cr
K_{3s}^{\rm SP-T}  =&  -4\sqrt{2}\alpha\beta_{\ell}\big[\im[\wilson{\rm T}\wilson[-*]{\rm P}+\wilson{\rm T5}\wilson[-*]{\rm S}]H_{1/2,1/2}^{\rm P}H_{-1/2,0,-}^{{\rm T},1/2}\cr
&+\im[\wilson{\rm T}\wilson[+*]{\rm S}+\wilson{\rm T5}\wilson[+*]{\rm P}]H_{1/2,1/2}^{\rm S}H_{-1/2,t,-}^{{\rm T},1/2} \big]  \, ,\cr
K_{4s}^{\rm SP-T}  =&  -4\sqrt{2}\alpha\beta_{\ell}\big[\real[\wilson{\rm T}\wilson[-*]{\rm S}+\wilson{\rm T5}\wilson[-*]{\rm P}]H_{1/2,1/2}^{\rm P}H_{-1/2,t,-}^{{\rm T},1/2}\nn\\
&+\real[\wilson{\rm T}\wilson[+*]{\rm P}+\wilson{\rm T5}\wilson[+*]{\rm S}]H_{1/2,1/2}^{\rm S}H_{-1/2,0,-}^{{\rm T},1/2} \big] \, .
\end{align}
The nonzero angular coefficients of  interference between  the vector (axial-vector) and electromagnetic  $\op{7(')}$ operators can be collected as follows
\begin{align}
K_{1c}^{\rm VA-Tq}  =&  -2\beta_{\ell}\big[\real[\wilson[+]{7}\wilson[-*]{10}]H_{-1/2,1/2,+}^{\rm A}H_{-1/2,1/2,+}^{\rm Tq}+\real[\wilson[-]{7}\wilson[+*]{10}]H_{-1/2,1/2,+}^{\rm V}H_{-1/2,1/2,+}^{\rm Tq5} \big]  \, ,\cr
K_{1cc}^{\rm VA-Tq} =&  2\real[\wilson[-]{7}\wilson[-*]{9}]H_{-1/2,1/2,+}^{\rm A}H_{-1/2,1/2,+}^{\rm Tq5}+2\real[\wilson[+]{7}\wilson[+*]{9}]H_{-1/2,1/2,+}^{\rm V}H_{-1/2,1/2,+}^{\rm Tq}  \cr
&+\frac{8m_{\ell}^2}{q^2} \big[\real[\wilson[-]{7}\wilson[-*]{9}]H_{1/2,1/2,0}^{\rm A}H_{1/2,1/2,0}^{\rm Tq5}+\real[\wilson[+]{7}\wilson[+*]{9}]H_{1/2,1/2,0}^{\rm V}H_{1/2,1/2,0}^{\rm Tq} \big] \,,\cr
K_{1ss}^{\rm VA-Tq} =&  2\real[\wilson[-]{7}\wilson[-*]{9}]H_{1/2,1/2,0}^{\rm A}H_{1/2,1/2,0}^{\rm Tq5}+\beta_{\ell+}^{2} \real[\wilson[-]{7}\wilson[-*]{9}]H_{-1/2,1/2,+}^{\rm A}H_{-1/2,1/2,+}^{\rm Tq5}  \cr
&+2\real[\wilson[+]{7}\wilson[+*]{9}]H_{1/2,1/2,0}^{\rm V}H_{1/2,1/2,0}^{\rm Tq}+\beta_{\ell+}^{2}\real[\wilson[+]{7}\wilson[+*]{9}]H_{-1/2,1/2,+}^{\rm V}H_{-1/2,1/2,+}^{\rm Tq}  \,,\cr
K_{2c}^{\rm VA-Tq}  =&  -2\alpha\beta_{\ell}\big[\real[\wilson[-]{7}\wilson[-*]{10}]H_{-1/2,1/2,+}^{\rm A}H_{-1/2,1/2,+}^{\rm Tq5}+\real[\wilson[+]{7}\wilson[+*]{10}]H_{-1/2,1/2,+}^{\rm V}H_{-1/2,1/2,+}^{\rm Tq} \big]  \, ,\cr
K_{2cc}^{\rm VA-Tq} =&  2\alpha\real[\wilson[-]{7}\wilson[+*]{9}]H_{-1/2,1/2,+}^{\rm V}H_{-1/2,1/2,+}^{\rm Tq5}+2\alpha\real[\wilson[+]{7}\wilson[-*]{9}]H_{-1/2,1/2,+}^{\rm A}H_{-1/2,1/2,+}^{\rm Tq}  \cr
&+\frac{8\alpha m_{\ell}^2}{q^2} \big[\real[\wilson[-]{7}\wilson[+*]{9}]H_{1/2,1/2,0}^{\rm V}H_{1/2,1/2,0}^{\rm Tq5}+\real[\wilson[+]{7}\wilson[-*]{9}]H_{1/2,1/2,0}^{\rm A}H_{1/2,1/2,0}^{\rm Tq} \big]   \,,\cr
K_{2ss}^{\rm VA-Tq} =&  2\alpha\real[\wilson[-]{7}\wilson[+*]{9}]H_{1/2,1/2,0}^{\rm V}H_{1/2,1/2,0}^{\rm Tq5}+ \alpha\beta_{\ell+}^{2} \real[\wilson[-]{7}\wilson[+*]{9}]H_{-1/2,1/2,+}^{\rm V}H_{-1/2,1/2,+}^{\rm Tq5} \cr
&+2\alpha\real[\wilson[+]{7}\wilson[-*]{9}]H_{1/2,1/2,0}^{\rm A}H_{1/2,1/2,0}^{\rm Tq}+\alpha\beta_{\ell+}^{2}\real[\wilson[+]{7}\wilson[-*]{9}]H_{-1/2,1/2,+}^{\rm A}H_{-1/2,1/2,+}^{\rm Tq}   \, ,\cr
K_{3sc}^{\rm VA-Tq} =&  \sqrt{2}\alpha\beta_{\ell}^2\big[\im[\wilson[-]{9}\wilson[-*]{7}](H_{1/2,1/2,0}^{\rm A}H_{-1/2,1/2,+}^{\rm Tq5}-H_{-1/2,1/2,+}^{\rm A}H_{1/2,1/2,0}^{\rm Tq5}) \cr 
&+\im[\wilson[+]{9}\wilson[+*]{7}](H_{-1/2,1/2,+}^{\rm V}H_{1/2,1/2,0}^{\rm Tq}-H_{1/2,1/2,0}^{\rm V}H_{-1/2,1/2,+}^{\rm Tq})\big]\, ,\cr
K_{3s}^{\rm VA-Tq}  =&  \sqrt{2}\alpha\beta_{\ell}\big[\im[\wilson[+]{7}\wilson[-*]{10}](H_{1/2,1/2,0}^{\rm A}H_{-1/2,1/2,+}^{\rm Tq}+H_{-1/2,1/2,+}^{\rm A}H_{1/2,1/2,0}^{\rm Tq}) \cr 
&-\im[\wilson[-]{7}\wilson[+*]{10}](H_{1/2,1/2,0}^{\rm V}H_{-1/2,1/2,+}^{\rm Tq5}+H_{-1/2,1/2,+}^{\rm V}H_{1/2,1/2,0}^{\rm Tq5})\big]\, ,\cr
K_{4sc}^{\rm VA-Tq} =&  \sqrt{2}\alpha\beta_{\ell}^2\big[\real[\wilson[+]{9}\wilson[-*]{7}](H_{-1/2,1/2,+}^{\rm V}H_{1/2,1/2,0}^{\rm Tq5}-H_{1/2,1/2,0}^{\rm V}H_{-1/2,1/2,+}^{\rm Tq5}) \cr 
&+\real[\wilson[-]{9}\wilson[+*]{7}](H_{1/2,1/2,0}^{\rm A}H_{-1/2,1/2,+}^{\rm Tq}-H_{-1/2,1/2,+}^{\rm A}H_{1/2,1/2,0}^{\rm Tq})\big] \,  ,\cr
K_{4s}^{\rm VA-Tq}  =&   \sqrt{2}\alpha\beta_{\ell}\big[ \real[\wilson[+]{7}\wilson[+*]{10}](H_{-1/2,1/2,+}^{\rm V}H_{1/2,1/2,0}^{\rm Tq}+H_{1/2,1/2,0}^{\rm V}H_{-1/2,1/2,+}^{\rm Tq})\cr 
&-\real[\wilson[-]{7}\wilson[-*]{10}](H_{1/2,1/2,0}^{\rm A}H_{-1/2,1/2,+}^{\rm Tq5}+H_{-1/2,1/2,+}^{\rm A}H_{1/2,1/2,0}^{\rm Tq5})\big] \, .
\end{align}
For the nonzero angular coefficients of interference between  the vector (axial-vector) and tensor operators, we can obtain
\begin{align}
K_{1c}^{\rm VA-T}  =&  \frac{16\beta_{\ell} m_\ell}{\sqrt{q^2}}\big[\real[\wilson{\rm T}\wilson[-*]{10}](-H_{-1/2,1/2,+}^{\rm A}H_{-1/2,t,-}^{{\rm T},1/2}+H_{1/2,1/2,t}^{\rm A}H_{1/2,+,-}^{{\rm T},1/2}) \cr
&+\real[\wilson{\rm T5}\wilson[+*]{10}](H_{-1/2,1/2,+}^{\rm V}H_{-1/2,0,-}^{{\rm T},1/2}+H_{1/2,1/2,t}^{\rm V}H_{1/2,t,0}^{{\rm T},1/2})\big] \, ,\cr
K_{1cc}^{\rm VA-T} =&  \frac{16 m_\ell}{\sqrt{q^2}}\big[\real[\wilson{\rm T5}\wilson[-*]{9}](-H_{-1/2,1/2,+}^{\rm A}H_{-1/2,0,-}^{{\rm T},1/2}+H_{1/2,1/2,0}^{\rm A}H_{1/2,+,-}^{{\rm T},1/2}) \cr
&+\real[\wilson{\rm T}\wilson[+*]{9}](H_{-1/2,1/2,+}^{\rm V}H_{-1/2,t,-}^{{\rm T},1/2}+H_{1/2,1/2,0}^{\rm V}H_{1/2,t,0}^{{\rm T},1/2})\big]  \, ,\cr
K_{1ss}^{\rm VA-T} =&   K_{1cc}^{\rm VA-T} \, ,\cr
K_{2c}^{\rm VA-T}  =&  \frac{16\alpha\beta_{\ell} m_\ell}{\sqrt{q^2}}\big[\real[\wilson{\rm T5}\wilson[-*]{10}](H_{-1/2,1/2,+}^{\rm A}H_{-1/2,0,-}^{{\rm T},1/2}+H_{1/2,1/2,t}^{\rm A}H_{1/2,t,0}^{{\rm T},1/2}) \cr
&+\real[\wilson{\rm T}\wilson[+*]{10}](-H_{-1/2,1/2,+}^{\rm V}H_{-1/2,t,-}^{{\rm T},1/2}+H_{1/2,1/2,t}^{\rm V}H_{1/2,+,-}^{{\rm T},1/2})\big]   \,,\cr
K_{2cc}^{\rm VA-T} =&  \frac{16\alpha m_\ell}{\sqrt{q^2}}\big[\real[\wilson{\rm T}\wilson[-*]{9}](H_{-1/2,1/2,+}^{\rm A}H_{-1/2,t,-}^{{\rm T},1/2}+H_{1/2,1/2,0}^{\rm A}H_{1/2,t,0}^{{\rm T},1/2}) \cr
&+\real[\wilson{\rm T5}\wilson[+*]{9}](-H_{-1/2,1/2,+}^{\rm V}H_{-1/2,0,-}^{{\rm T},1/2}+H_{1/2,1/2,0}^{\rm V}H_{1/2,+,-}^{{\rm T},1/2})\big]  \, ,\cr
K_{2ss}^{\rm VA-T} =& K_{2cc}^{\rm VA-T}   \, ,\cr
K_{3s}^{\rm VA-T}  =&  \frac{8\sqrt{2}\alpha\beta_{\ell} m_\ell}{\sqrt{q^2}}\big[\im[\wilson{\rm T}\wilson[-*]{10}](-H_{1/2,1/2,t}^{\rm A}H_{-1/2,0,-}^{{\rm T},1/2}+H_{1/2,1/2,0}^{\rm A}H_{-1/2,t,-}^{{\rm T},1/2}\cr
&+H_{-1/2,1/2,+}^{\rm A}H_{1/2,t,0}^{{\rm T},1/2})+\im[\wilson{\rm T5}\wilson[+*]{10}](H_{1/2,1/2,0}^{\rm V}H_{-1/2,0,-}^{{\rm T},1/2}\cr
&-H_{1/2,1/2,t}^{\rm V}H_{-1/2,t,-}^{{\rm T},1/2}-H_{-1/2,1/2,+}^{\rm V}H_{1/2,+,-}^{{\rm T},1/2})\big]  \, ,\cr
K_{4s}^{\rm VA-T}  =&  \frac{8\sqrt{2}\alpha\beta_{\ell} m_\ell}{\sqrt{q^2}}\big[\im[\wilson{\rm T5}\wilson[-*]{10}](H_{1/2,1/2,0}^{\rm A}H_{-1/2,0,-}^{{\rm T},1/2}-H_{1/2,1/2,t}^{\rm A}H_{-1/2,t,-}^{{\rm T},1/2}\cr
&-H_{-1/2,1/2,+}^{\rm A}H_{1/2,+,-}^{{\rm T},1/2}) +\im[\wilson{\rm T}\wilson[+*]{10}](-H_{1/2,1/2,t}^{\rm V}H_{-1/2,0,-}^{{\rm T},1/2}\cr
&+H_{1/2,1/2,0}^{\rm V}H_{-1/2,t,-}^{{\rm T},1/2}+H_{-1/2,1/2,+}^{\rm V}H_{1/2,t,0}^{{\rm T},1/2})\big]  \,.
\end{align}
For the nonzero angular coefficients of  interference between  the tensor and electromagnetic operators, we can get
\begin{align}
K_{1cc}^{\rm Tq-T} =&  \frac{16 m_\ell}{\sqrt{q^2}}\big[\real[\wilson{\rm T5}\wilson[-*]{7}](-H_{-1/2,1/2,+}^{\rm Tq5}H_{-1/2,0,-}^{{\rm T},1/2}+H_{1/2,1/2,0}^{\rm Tq5}H_{1/2,+,-}^{{\rm T},1/2}) \cr
&+\real[\wilson{\rm T}\wilson[+*]{7}](H_{-1/2,1/2,+}^{\rm Tq}H_{-1/2,t,-}^{{\rm T},1/2}+H_{1/2,1/2,0}^{\rm Tq}H_{1/2,t,0}^{{\rm T},1/2})\big]  \, ,\cr
K_{1ss}^{\rm Tq-T} =&  K_{1cc}^{\rm Tq-T}  \, ,\cr
K_{2cc}^{\rm Tq-T} =&  \frac{16\alpha m_\ell}{\sqrt{q^2}}\big[\real[\wilson{\rm T}\wilson[-*]{7}](H_{-1/2,1/2,+}^{\rm Tq5}H_{-1/2,t,-}^{{\rm T},1/2}+H_{1/2,1/2,0}^{\rm Tq5}H_{1/2,t,0}^{{\rm T},1/2}) \cr
&+\real[\wilson{\rm T5}\wilson[+*]{7}](-H_{-1/2,1/2,+}^{\rm Tq}H_{-1/2,0,-}^{{\rm T},1/2}+H_{1/2,1/2,0}^{\rm Tq}H_{1/2,+,-}^{{\rm T},1/2})\big]  \, ,\cr
K_{2ss}^{\rm Tq-T} =& K_{2cc}^{\rm Tq-T}    \, .
\end{align}

Compared with the current research progress, our calculations include the contributions of all six dimensional operators in $b \to s \ell ^+  \ell ^ - $ transitions, and the contributions of all possible interference between these operators are taken into account, without neglecting lepton mass.  
Assuming all leptons are massless, the summation of $K_{\{\cdots\}}^{\rm VA}+K_{\{\cdots\}}^{\rm Tq}+K_{\{\cdots\}}^{\rm VA-Tq}$ are completely consistent with Ref.~\cite{Boer:2014kda},  
and the reliability of our calculation is verified from the side. In the attachment of ArXiv version,  a Mathematica program is provided to prove the consistency with results of Ref.~\cite{Boer:2014kda}, and all expressions in this paper are also available in Mathematica.
\subsection{Observables}\label{sec:obsdefLamb2Lam}
Usually, one can construct observables through weighted angular integrals of the differential distributions Eq.~(\ref{eq:angular-distribution})~\cite{Boer:2014kda}~,
\begin{align}
    X(q^2) \equiv \int \frac{\dd^4 \Gamma}{\dd q^2\, \dd\cos\theta_\ell\, \dd\cos\theta_\Lambda\, \dd \phi}
    \, \omega_X(q^2,\cos\theta_\ell,\cos\theta_\Lambda,\phi) \, \dd \cos\theta_\ell \, \dd \cos\theta_\Lambda \, \dd \phi \,.
\end{align}
By integrating over $\theta_\ell\in[0,\pi]$, $\theta_\Lambda\in[0,\pi]$, $\phi\in[0,2\pi]$, the observables we consider are as follows.
\begin{itemize}
 \item The differential decay rate and differential branching fraction are 
 \begin{align}
    \frac{\dd \Gamma}{\dd q^2}  = 2 \overline{K}_{1ss} + \overline{K}_{1cc}\, ,\quad \frac{\dd\mB}{\dd q^2}=\tau_{\Lambda_b}\frac{\dd \Gamma}{\dd q^2}\, , \quad~\omega_X\equiv 1\, .
\end{align}

\item
The longitudinal polarization fraction and corresponding weighting factor are defined as
\begin{align}
 F_{\rm L}  &= \frac{2 \overline{K}_{1ss} - \overline{K}_{1cc}}{2 \overline{K}_{1ss} + \overline{K}_{1cc}} \,,
&\omega_{F_{\rm L}} &= \frac{2-5 \cos^2\theta_\ell}{d\Gamma/dq^2} \,.
\end{align}
%

\item The lepton-side forward-backward asymmetry $A_{\rm FB}^\ell$ is defined as follows
\begin{align}
    A^\ell_\text{FB} & =
    \frac{3}{2} \, \frac{\overline{K}_{1c}}{2 \overline{K}_{1ss} + \overline{K}_{1cc}}    \,,
    &    \omega_{A^\ell_\text{FB}}          & = \frac{{\rm sign} [\cos\theta_\ell]}{d\Gamma/dq^2}
    \,.
\end{align}
\item  The hadron-side forward-backward asymmetry $A_{\rm FB}^\Lambda$ is defined as follows
\begin{align} 
  A^\Lambda_\text{FB}
  & = \frac{1}{2} \, \frac{2 \overline{K}_{2ss} + \overline{K}_{2cc}}{2 \overline{K}_{1ss} + \overline{K}_{1cc}}  \,,  &
  \omega_{A^\Lambda_\text{FB}}
  & = \frac{{\rm sign} [\cos\theta_\Lambda]}{d\Gamma/dq^2} \,.
\end{align}
\item The lepton-hadron-side  forward-backward asymmetry $A_{\rm FB}^{\ell\Lambda}$ is defined as follows
\begin{align}
    A^{\ell\Lambda}_\text{FB} & = \frac{3}{4} \, \frac{\overline{K}_{2c}}{2 \overline{K}_{1ss} +\overline{K}_{1cc}}\,,
    &    \omega_{A^{\ell\Lambda}_\text{FB}} & = \frac{{\rm sign} [\cos\theta_\ell\,\cos\theta_\Lambda]}{d\Gamma/dq^2}\,.
\end{align}

\item The lepton flavour universality ratio is written as
\begin{align}
R_{\mu /e}=\frac{\displaystyle\int_{q^2_{\rm min}}^{q^2_{\rm max}} \frac{d 
\Gamma(\Lambda_{b} 
\rightarrow \Lambda\mu^+\mu^-)}{dq^2}dq^2}{\displaystyle\int_{q^2_{\rm 
min}}^{q^2_{\rm 
max}} 
\frac{d \Gamma(\Lambda_{b} \rightarrow \Lambda e^+e^-)}{dq^2}dq^2}\,.
\end{align}

\item In order to compare with the experimental data of Ref.~\cite{Aaij:2015xza}, we also define the normalized angular observables
\begin{align}
\langle \dd \mathcal{B}/\dd q^2\rangle_{[q_{\rm min}^2,q_{\rm max}^2]}=\frac{\int_{q_{\rm min}^2}^{q_{\rm max}^2}(\dd \mathcal{B}/\dd q^2)\dd q^2}{q_{\rm max}^2-q_{\rm min}^2}\,,
\end{align}
\begin{align}
\langle A\rangle_{[q_{\rm min}^2,q_{\rm max}^2]}=\frac{\int_{q_{\rm min}^2}^{q_{\rm max}^2}A\dd q^2}{\int_{q_{\rm min}^2}^{q_{\rm max}^2}(\dd \Gamma/\dd q^2)\dd q^2}\,,
\end{align}
where $A$ can denote $F_{\rm L}$, $A_{\rm FB}^{\ell,~\Lambda,~\ell\Lambda}$ and $K_{2ss,~2cc,~4s,~4sc}$, as shown in Tab. \ref{tab:binnedobs}. 
\end{itemize}
\section{Numerical Analysis}\label{sec:numLamb2Lam}
\begin{table}
\caption{\label{tab:inputsLamb2Lam} List of input parameters.}
\begin{center}
\begin{tabular}{cr|cr}
\toprule \toprule
Inputs & Values & Inputs & Values\\
\hline\noalign{\smallskip}
$ \alpha_e(m_b) $  & $1/127.925(16)$ ~\cite{PDG:2018} & $|V_{tb}V_{ts}^\ast|$  & $0.0401 \pm 0.0010$ ~\cite{PDG:2018,web:UTfit}  \\ 
	$m_c(\overline{\text{MS}})$ & $1.27\pm 0.03$ GeV ~\cite{PDG:2018} & $m_{\Lambda_b}$ & 5.619 GeV ~\cite{PDG:2018} \\ 
	$\mu_b$ &  $4.2$ GeV ~\cite{Detmold:2016pkz} & $m_\Lambda $ & $1.115$ GeV ~\cite{PDG:2018} \\
	$m_b(\overline{\text{MS}})$ & $4.18\pm 0.03$ GeV ~\cite{PDG:2018} & $\tau_{\Lambda_b}$ & $(1.470\pm 0.010)\times 10^{-12} s$ ~\cite{PDG:2018} \\  
	$m_{b}({\rm pole})$ & $4.78\pm 0.06$ GeV ~\cite{PDG:2018}  & $m_{B^0}$ & $5.279$ GeV ~\cite{PDG:2018} \\ 
	$\alpha_s(m_b)$ & $0.2233$ ~\cite{Detmold:2016pkz}  & $m_K$ & $0.494$ GeV ~\cite{PDG:2018} \\ 
	$m_{c}({\rm pole})$ & $1.67\pm 0.07$ ~\cite{PDG:2018} & $\alpha_\Lambda$ & $0.642\pm0.013$ ~\cite{PDG:2018}\\
	$\alpha_s(M_Z)$ & $0.1181\pm 0.0011$ ~\cite{PDG:2018} & $m_c(m_c)$ & $1.28\pm 0.03$ GeV ~\cite{PDG:2018}\\
\bottomrule \bottomrule
\end{tabular}
\end{center}
\end{table}

For comparison with the experimental data, we study the numerical results of the $\Lambda_b\to\Lambda(\to N\pi)\ell^+\ell^-$~observables described above. Numerical analysis would require a series of input parameters, such as quark mass, Wilson coefficients, form factors and so on. 
In Tab.~\ref{tab:inputsLamb2Lam}, we collect the input parameters involved in our numerical analysis. They include electromagnetic and strong coupling constants, quark and hadron masses and so on. 

\subsection{ ~$\Lambda_b\to \Lambda$~form factors}\label{sec:FFLamb2Lam}
In addition to the above-mentioned input parameters, the form factors are also  important inputs in $\Lambda_b\to \Lambda\ell^+\ell^-$ decay. For the $\Lambda_b\to \Lambda$ transition form factors, we adopt the latest Lattice QCD results~\cite{Detmold:2016pkz} with  2+1 flavour dynamics. Explicit expressions of all the relevant transition form factors are in Sec. 2 of Ref.~\cite{Detmold:2016pkz}. 

The $\Lambda_b\rightarrow\Lambda$ hadronic matrix elements can be written in terms of ten $q^2$ dependent helicity  form factors $F_{0,+,\perp},G_{0,+,\perp},h_{+,\perp},\widetilde{h}_{+,\perp}$ in Refs.~\cite{Detmold:2015aaa,Datta:2017aue,Detmold:2016pkz}. Following Ref.~\cite{Detmold:2015aaa}, the lattice calculations are fitted to two BCL $z$-parameterizations. In the so called `` nominal-order " fit, a form factor $f$ reduces to the form
\begin{eqnarray}
f(q^2) &=& \frac{1}{1-q^2/(m_{\rm pole}^f)^2} \big[ a_0^f + a_1^f\:z^f(q^2)  \big]\, , \label{eq:nominalfitphys}
\end{eqnarray}
while in the `` higher-order " fit, a form factor $f$  is given by
\begin{eqnarray}
f_{\rm HO}(q^2) &=& \frac{1}{1-q^2/(m_{\rm pole}^f)^2} \big[ a_{0,{\rm HO}}^f + a_{1,{\rm HO}}^f\:z^f(q^2) + a_{2,{\rm HO}}^f\:[z^f(q^2)]^2  \big]\, , \label{eq:HOfitphys}
\end{eqnarray}
where~$$z^f(q^2) = \frac{\sqrt{t_+^f-q^2}-\sqrt{t_+^f-t_0}}{\sqrt{t_+^f-q^2}+\sqrt{t_+^f-t_0}}\, ,
\quad t_0 = (m_{\Lambda_b} - m_{\Lambda})^2\, ,
\quad t_+^f = (m_{\rm pole}^f)^2\,.$$

The values of the fit parameters and all the pole masses are taken from Ref.~\cite{Detmold:2016pkz}.
The method to estimate the central values, the statistical and systematic uncertainties for any observable $\mathcal{O}$ are as follows~\cite{Detmold:2016pkz}.
\begin{itemize}
\item The central value and uncertainty computed using the nominal-order fit~Eq.~(\ref{eq:nominalfitphys}) are denoted as
$$
\mathcal{O}\,,\sigma_{\rm O}\,.
$$
\item The central value and uncertainty obtained using the higher-order fit~Eq.~(\ref{eq:HOfitphys}) are denoted as
$$
\mathcal{O}_{\rm HO}\,,\sigma_{\rm O\,,HO}\,.
$$
\item Then, the final results are written as
$$
\mathcal{O}\pm\sigma_{\rm O,stat}\pm\sigma_{\rm O,syst}\,,
$$
where $\mathcal{O}\,,\sigma_{\rm O\,,stat}$ and $\sigma_{\rm O\,,syst}$ denote the central value, statistical and systematic uncertainty, 
\begin{align*}
\sigma_{\rm O,stat}=\sigma_{\rm O}\, ,
\quad\sigma_{\rm O,syst}={\rm max}\{ |\mathcal{O}_{\rm HO}-\mathcal{O}|\, ,~H(\sigma_{\rm O,HO}-\sigma_{\rm O})\sqrt{|\sigma_{\rm O,HO}^2-\sigma_{\rm O}^2|} \}\, ,
\end{align*}
and $H$ is the Heaviside step function.
\end{itemize}


In brief, the central value and statistical uncertainty of observables are obtained from the nominal-order fit, and the systematic uncertainty is given by the larger of nominal and higher order fit. For more accurate results, we consider all the correlations among the fit parameters in our numerical analysis. 

\subsection{Results within the SM}
\begin{figure}[t]
  \begin{center}
  \includegraphics[width=0.45\linewidth]{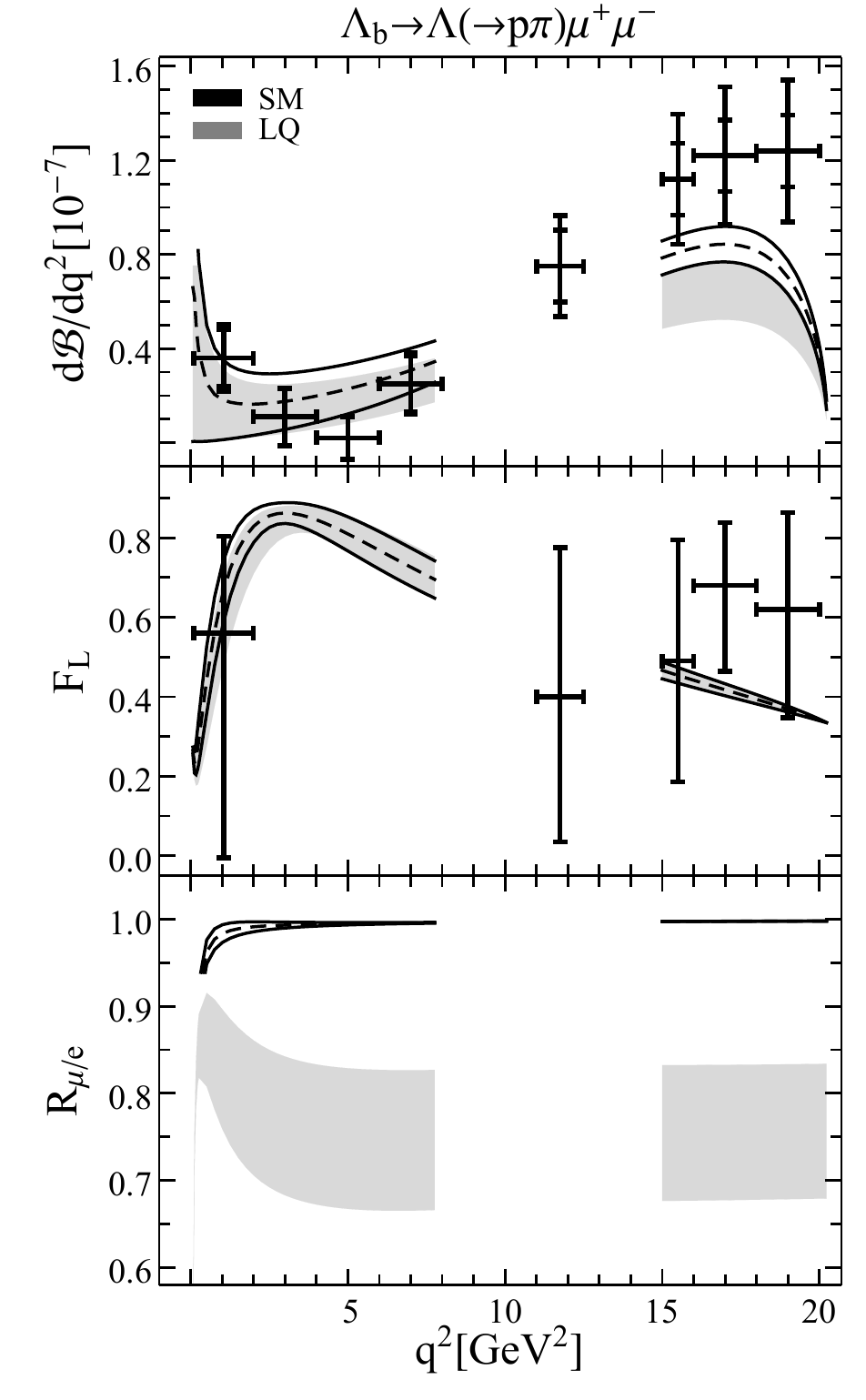}
  \includegraphics[width=0.45\linewidth]{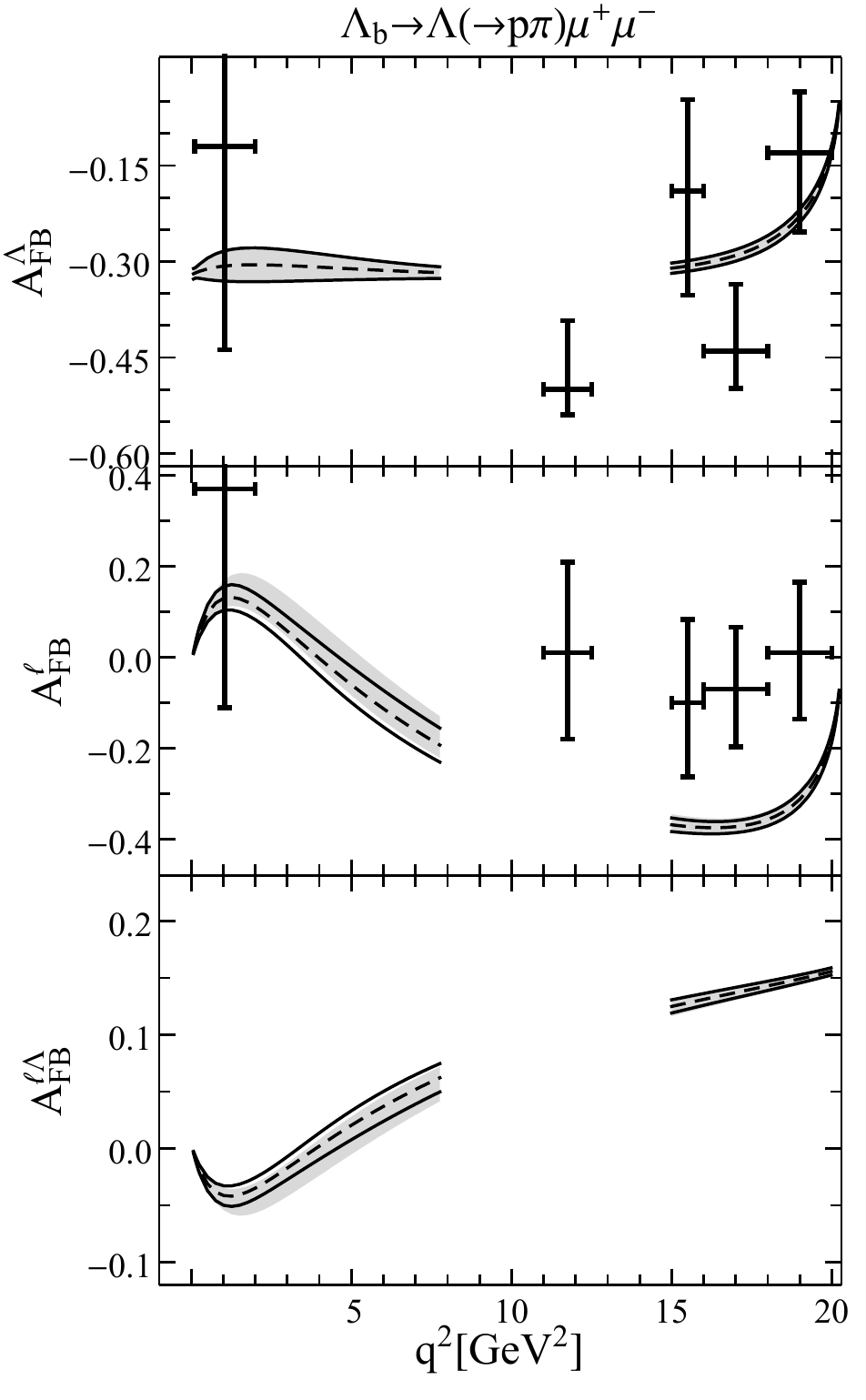}
  \caption{\label{fig:LamB2LamS1pS3}The $q^2$ distributions of the observables in ~$\Lambda_b\to\Lambda(\to p\pi)\mu^+\mu^-$~decay. The black curves (gray band) indicate the SM ($S_1+S_3$~LQ) central values with $1\sigma$ theoretical uncertainty. The corresponding experimental data from LHCb ~\cite{Aaij:2015xza}, where available, are represented by the error bars.}
  \end{center}
\end{figure}

\begin{table}[t]
\begin{center}
\caption{\label{tab:binnedobs}SM predictions for the differential branching fraction~(\, in units of~$10^{-7}~{\rm GeV}^{-2}$~) and angular observables. The first column specifies the bin ranges~$[q^2_{\rm min},~q^2_{\rm max}]$~in units of~${\rm GeV}^2$.}
\vspace{0.5em}
\begin{tabular}{llllll}
\toprule \toprule
& $\phantom{-}\langle \mathrm{d}\mathcal{B}/\mathrm{d} q^2\rangle$ & $\phantom{-}\langle F_{\rm L} \rangle $ & $\phantom{-}\langle A_{\rm FB}^\ell \rangle $ & $\phantom{-}\langle A_{\rm FB}^\Lambda \rangle $ &  $\phantom{-}\langle A_{\rm FB}^{\ell\Lambda} \rangle $   \\
\hline\noalign{\smallskip}
$[0.1, 2]$    & $\phantom{-}0.226(211)$  & $\phantom{-}0.542(80)$  & $\phantom{-}0.098(16)$  & $-0.310(18)$   & $-0.031(5)$   \\ 
$[2, 4]$      & $\phantom{-}0.177(120)$  & $\phantom{-}0.854(27)$  & $\phantom{-}0.055(31)$  & $-0.307(24)$   & $-0.016(10)$  \\ 
$[4, 6]$      & $\phantom{-}0.229(106)$  & $\phantom{-}0.807(43)$  & $-0.062(39)$            & $-0.311(17)$   & $\phantom{-}0.021(13)$    \\ 
$[6, 8]$      & $\phantom{-}0.309(93)$   & $\phantom{-}0.724(48)$  & $-0.162(39)$            & $-0.316(11)$   & $\phantom{-}0.052(13)$     \\ 
$[1.1, 6]$    & $\phantom{-}0.197(119)$  & $\phantom{-}0.818(32)$  & $\phantom{-}0.010(30)$  & $-0.309(21)$   & $-0.002(10)$   \\ 
$[15, 16]$    & $\phantom{-}0.808(73)$   & $\phantom{-}0.454(20)$  & $-0.372(14)$            & $-0.307(8)$   & $\phantom{-}0.128(6)$   \\ 
$[16, 18]$    & $\phantom{-}0.836(75)$   & $\phantom{-}0.418(15)$  & $-0.370(13)$            & $-0.289(8)$   & $\phantom{-}0.137(5)$    \\ 
$[18, 20]$    & $\phantom{-}0.669(67)$   & $\phantom{-}0.371(8)$   & $-0.307(15)$            & $-0.227(10)$   & $\phantom{-}0.149(4)$   \\ 
$[15, 20]$    & $\phantom{-}0.764(69)$   & $\phantom{-}0.409(13)$  & $-0.349(13)$            & $-0.271(9)$   & $\phantom{-}0.139(4)$    \\
\hline\hline\noalign{\smallskip}
& $\phantom{-}\langle K_{2ss} \rangle $ & $\phantom{-}\langle K_{2cc} \rangle $ & $\phantom{-}\langle K_{4s} \rangle $ &  $\phantom{-}\langle {K}_{4sc} \rangle $ & \\
\hline\noalign{\smallskip}
$[0.1, 2]$    & $-0.237(20)$   & $-0.146(26)$  & $\phantom{-}0.009(23)$     & $-0.022(22)$ &\\ 
$[2, 4]$      & $-0.284(23)$   & $-0.045(9)$   & $-0.030(36)$               & $-0.012(31)$ &\\ 
$[4, 6]$      & $-0.281(15)$   & $-0.061(14)$  & $-0.037(44)$               & $-0.00047(3096)$ &\\ 
$[6, 8]$      & $-0.272(10)$   & $-0.088(15)$  & $-0.030(38)$               & $\phantom{-}0.007(27)$ &\\ 
$[1.1, 6]$    & $-0.280(20)$   & $-0.057(10)$  & $-0.030(35)$               & $-0.008(29)$ &\\ 
$[15, 16]$    & $-0.225(7)$    & $-0.164(7)$   & $\phantom{-}0.060(13)$     & $\phantom{-}0.021(8)$ &\\ 
$[16, 18]$    & $-0.208(7)$    & $-0.162(7)$   & $\phantom{-}0.089(10)$     & $\phantom{-}0.021(6)$ &\\ 
$[18, 20]$    & $-0.159(7)$    & $-0.134(7)$   & $\phantom{-}0.145(8)$      & $\phantom{-}0.017(4)$ &\\ 
$[15, 20]$    & $-0.194(7)$    & $-0.153(6)$   & $\phantom{-}0.103(10)$     & $\phantom{-}0.019(5)$ &\\
\bottomrule \bottomrule
\end{tabular}
\end{center}

\end{table}
\begin{table}
\caption{\label{tab:binnedsmobs}Comparison between the results of SM predictions, $S_1+S_3$ Leptoquark (LQ) \cite{Crivellin:2017zlb,Yan:2019hpm} and LHCb data for the process~$\Lambda_b \to \Lambda\, \mu^+ \mu^-$.
}
\begin{center}
\begin{tabular}{c|rrr}
\toprule \toprule
  & SM & LQ & LHCb~\cite{Aaij:2015xza}\\
\hline\noalign{\smallskip}
$ \langle{\rm d}\mathcal{B}/{\rm d}q^2\rangle_{[1,\,6]} \times 10^{7}$  &$0.197\pm 0.119$ &$[0.053,\,0.267]$ & $0.09_{-0.006\,-0.05}^{+0.01\,+0.01}\pm 0.02$  \\ 
$ \langle{\rm d}\mathcal{B}/{\rm d}q^2\rangle_{[15,\,20]}\times 10^{7} $  &$0.764\pm 0.069$& $[0.472,\,0.695]$ & $1.20_{-0.09\,-0.04}^{+0.09\,+0.02}\pm 0.25$  \\
	$\langle F_{\rm L}\rangle_{[15,\,20]}$ &  $0.409\pm0.013$ &$[0.397,0.424]$ & $0.61_{-0.14}^{+0.11}\pm0.03$\\ 
	$\langle A_{\rm FB}^{\ell}\rangle_{[15,\,20]}$ & $-0.349\pm0.013$ &$[-0.360,\,-0.330]$ & $-0.05\pm 0.09\pm0.003$ \\
	$\langle A_{\rm FB}^{\Lambda}\rangle_{[15,\,20]}$ & $-0.271\pm0.009$ &$[-0.280,\,-0.262]$ &$-0.29\pm 0.07\pm 0.003$  \\ 
\bottomrule \bottomrule
\end{tabular}
\end{center}
\end{table}
Using the theoretical framework described in Sec.~\ref{sec:obsdefLamb2Lam}, the SM predictions for $\Lambda_b \to \Lambda\mu^+ \mu^-$ decay are presented in Tab.~(\ref{tab:binnedobs},~\ref{tab:binnedsmobs}) and Fig.~\ref{fig:LamB2LamS1pS3}. To obtain the theoretical uncertainties, we vary each input parameters within their respective $1\sigma$ range and add each individual uncertainty in quadrature. For the uncertainties of transition form factors, the correlations among the fit parameters have been taken into account  \cite{Yan:2019hpm}.  In particular, for the $\Lambda_b \to \Lambda \mu^+ \mu^-$ decay, we follow the treatment of Ref.~\cite{Detmold:2016pkz} to obtain the statistical and systematic uncertainties induced by the $\Lambda_b \to \Lambda$ transition form factors. 

From Tab.~\ref{tab:binnedsmobs}, the SM prediction of lepton-side forward-backward asymmetry exceeds the LHCb measurements  by $3.3~\sigma$, while the hadronic side forward-backward asymmetry is only $0.3~\sigma$ away from the measurements accordingly. The predictions of branching fraction $\mathcal{B}$ and longitudinal polarization fraction $F_{\rm L}$ deviate from the LHCb measurements  by $1.6~\sigma$ and $1.8~\sigma$ in the bin $[15,\,20]\GeV^2$ respectively. The deviation of branching fraction from the measured value is $0.9\sigma$ in the bin $[1.1,\,6]\GeV^2$, but this does not mean it is in good agreement with the experiment measurements. As the theoretical uncertainty  is significantly larger in low $q^2$ region, compared with the high $q^2$ region. One of the possible reason of the small deviation is that the form factors are not very accurately calculated in this region. These discrepancies may provide a hint of NP.
\begin{figure}[t]
  \begin{center}
  \includegraphics[width=0.45\linewidth]{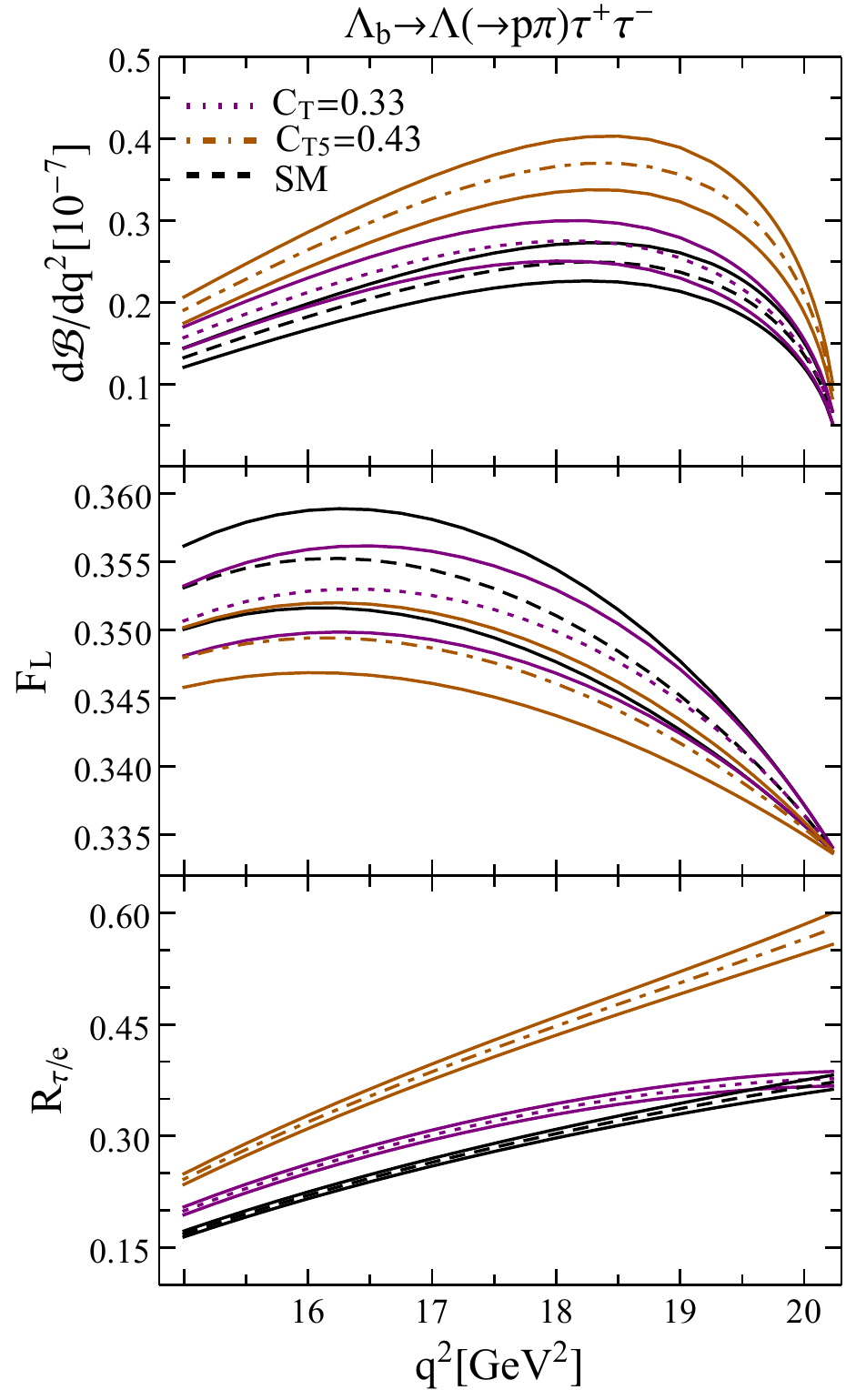}
  \includegraphics[width=0.45\linewidth]{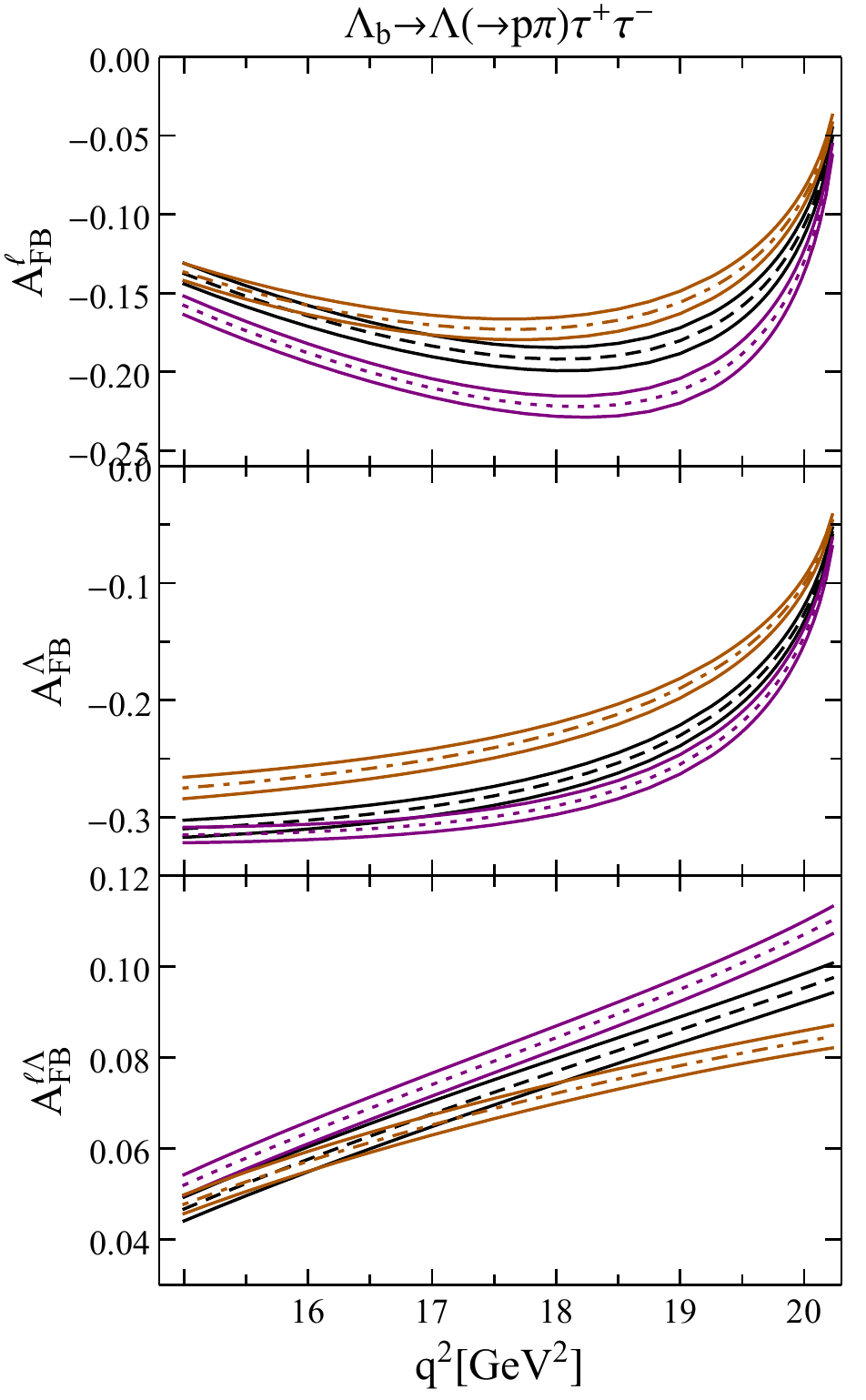}
  \caption{\label{fig:LamB2LamMI}The $q^2$distributions of the observables in~$\Lambda_b\to\Lambda(\to p\pi)\tau^+\tau^-$~decay. The black, purple and brown curves indicate the results of  three scenarios: SM, $\wilson{\rm T}=0.33$ and $\wilson{\rm T5}=0.43$ (central values: dotted line, theoretical uncertainties: solid line). }
  \end{center}
\end{figure}
\subsection{Model independent analysis of $\Lambda_b\to \Lambda(\to N \pi)\tau^+\tau^-$}\label{sec:numMILamb2Lam}
We have calculated contributions of the ~$\op{\rm T},~\op{\rm T5}$ without neglecting the lepton mass for the first time in this work, so we perform model independent NP analysis in this section. Following Ref.~\cite{Beaujean:2015gba}, which has restricted on tensor and scalar couplings from $B\to K\bar{\mu}\mu$ and $B_s\to \bar{\mu}\mu$ decay, we consider the following two scenarios, $\wilson{\rm T}=0.33$ and $\wilson{\rm T5}=0.43$. 
The $q^2$ distributions of the branching fraction $\mathcal B$, the LFU ratio $R_{\tau/e}$, 
the longitudinal polarization fraction $F_{\rm L}$, as well as the three kinds of forward-backward asymmetry $A_{\rm FB}^{\ell,\,\Lambda,\,\ell\Lambda}$ are shown in Fig.~\ref{fig:LamB2LamMI}. 
The bands of figures correspond to the uncertainties of form factors and inputs in Tab.~\ref{tab:inputsLamb2Lam}. 
From the figure, we find out that the differential branching fraction $\mathcal{B}$ and the LFU ratio $R_{\tau/e}$ get increase for both scenarios,  especially the enhancement effect of $\op{\rm T5}$ is more significant. However,  the longitudinal polarization fraction $F_{\rm L}$ is slight abate relative to the SM predictions.  More interestingly, $\op{\rm T}$ and $\op{\rm T5}$ have the opposite effect on the three kinds of forward-backward asymmetry as shown in the Fig.~\ref{fig:LamB2LamMI}. It can be deduced that the future measurements of  forward-backward asymmetry can provide ways of distinguishing $\op{\rm T}$ and $\op{\rm T5}$ operators.  In the future, more precise measurements of these distributions are important to confirm the existence of possible NP model in the $b\to s\ell^+\ell^-$ transitions. 
Through the above analysis, the possible NP effects of operators~$\op{\rm T},~\op{\rm T5}$~can not be ignored for $b\to s\ell^+\ell^-$ transitions. 
\subsection{Results in the $S_1+S_3$~LQ}
\label{sec:numS1S3Lamb2Lam}
In this section, we revisit the $S_1+S_3$ LQ model proposed in Ref.~\cite{Crivellin:2017zlb}, in which two scalar LQ, one being $SU(2)_L$ singlet and the other $SU(2)_L$ triplet, are introduced simultaneously. 
The LQ contributions to $b \to s \ell_i^+ \ell_j^-$ transitions are as follows \cite{Crivellin:2017zlb} 
\begin{align}
\mathcal{C}_9^{\NP,ij} =  - \mathcal{C}_{10}^{\NP,ij} = \frac{-\sqrt 2 }{2 G_{F} V_{tb}V_{ts}^*} \frac{\pi }{\alpha_e }\frac{1}{M^2}\lambda _{3j}^{L}\lambda _{2i}^{L*}\, .
\end{align}
In the model-independent approach, the current $b\to s \mu^+ \mu^-$ anomalies can be explained by contributions of $\mathcal{C}_{9}^{\NP,22} =  - \mathcal{C}_{10}^{\NP,22}$, with the allowed range given by~\cite{Altmannshofer:2015sma,Descotes-Genon:2015uva,Hurth:2016fbr,Angelescu:2018tyl}
\begin{align}
-0.91 \, (-0.71) \leq \mathcal{C}_{9}^{\NP,22}=-\mathcal{C}_{10}^{\NP,22} \leq  -0.18 \, (-0.35)\,,
\end{align} 
at the $2\sigma$~($1\sigma$) level, which provides in turn a constraint on $\lambda_{22}^{L*}\lambda_{32}^L$.
Following Ref.~\cite{Yan:2019hpm},  the allowed ranges of NP parameters $\lambda _{22}^{L*}\lambda _{32}^L$ at $1\sigma$ level are obtained 
\begin{align}\label{eq:lam22lam32}
0.549\times 10^{-3}<\lambda _{22}^{L*}\lambda _{32}^L <1.115\times 10^{-3} \, .
\end{align}

In the following discussion, we use the range of ~$\lambda _{22}^{L*}\lambda _{32}^L$~to obtain the $q^2$ distributions of the observables as shown in Fig.~\ref{fig:LamB2LamS1pS3}. 
For the $\Lambda_b \to \Lambda(\to N\pi) \mu^+ \mu^-$ decay,  the $q^2$ distributions of the branching fraction $\mathcal B$ is largely decreased by the LQ effects,  especially at high $q^2$ region. More importantly, the differential ratio $R_{\mu/e}(q^2)$ at whole $q^2$ region shows significant differences in the SM and LQ scenario.  The large difference between the SM and LQ predictions 
provide a testable signature of the LQ effects.  Measurements of the differential ratios $R_{\mu/e}(q^2)$ are crucial to confirm the LFU violation  anomaly and to test the $S_1+S_3$ LQ model.  The LQ scenario effect has less influence on the fraction of longitudinal polarization and three kinds of forward-backward asymmetry, because the numerator of observables are canceled out greatly by the denominator in  LQ scenario.  With future accurate measurements  of $\Lambda_b \to \Lambda(\to N\pi)\mu^+ \mu^-$, we expect that this prediction could provide helpful information about the LQ effects. 

\section{Conclusions}\label{sec:conLamb2Lam}
In this work, we have investigated the full angular distribution of the $\Lambda_b\to \Lambda(\to N \pi)\ell^+\ell^-$ decay, where the leptons are massive and the $\Lambda_b$ is unpolarized. We  compute the angular distributions within the effective Hamiltonian approach, which includes the scalar, pseudo-scalar, vector, axial-vector and tensor operators under the narrow width approximation. In particular, our calculations include the contribution of tensor operators  for the first time without neglecting the lepton mass. The four body angular distributions are expressed in terms of 10 angular coefficients which consist of helicity amplitudes. 
We have checked that our result are completely consistent with Ref.~\cite{Boer:2014kda}, in the case of massless lepton. This proves the reliability of our analytical calculation. For our numerical analysis, we compare the SM predictions with the current LHCb results, and significant deviation have been found between them. 
We study the NP effects of the LQ model and obtain the $q^2$ distributions of observables accordingly. 
Due to  the large uncertainties of the form factors and the current experimental measurement, the $S_1+S_3$ LQ model can not be tested very well at present. 
In addition, we demonstrate  the sensitivity of various angular observables to tensor operators contributions. Utilizing the operator sensitivity analysis, we find out that the  potential NP effects of operators~$\op{\rm T},~\op{\rm T5}$~can not be ignored in $b\to s\ell^+\ell^-$ transitions. 

The more precise experimental measurements of various angular observables in $b \to s \ell^+ \ell^-$ transitions can help to confirm possible NP explanations of the $R_{K^{(*)}}$ anomalies and to distinguish among the various NP candidates. Along with the experimental progresses of the SuperKEKB~\cite{Kou:2018nap} and  HL-LHC~\cite{Cerri:2018ypt}, our predictions of the observables can be further explored in future.

\section*{Acknowledgements}
Dr.~Xing-Bo Yuan, Chao-Jie Fan, Jun-Kang He and Cong Wang help me to overcome many hurdles in this work. 
I would like to thank Xin-Qiang Li, Philipp B\"oer, Thorsten Feldmann, Danny van Dyk and Diganta Das for their fruitful discussions. This work is supported by the National Natural Science Foundation of China under Grant Nos. 11775092, 11521064, 11435003, and 11805077. XY is also supported in part by the startup research funding from CCNU.

\begin{appendix}
\section{Kinematic conventions}\label{sec:kin}
The four-body kinematics relationships are taken as follows, 
\begin{align*}
\Lambda_b(p,s_{\Lambda_b})\to& \Lambda(k,s_{\Lambda})\ell^{+}(q_{1},s_{\ell^{+}})\ell^{-}(q_{2},s_{\ell^{-}})\, ,\nn\\
\Lambda(k,s_{\Lambda})\to& N(k_{1},s_{N})\pi(k_{2})\, ,\quad (N\pi={p\pi^{-},n\pi^{0}})\, ,\cr
k=&k_1+k_2\, ,\quad q=q_1+q_2\, ,\cr
p^\mu =(m_{\Lambda_{b}},0,0,0)\, ,&
\qquad
k^\mu =(E_{\Lambda},0,0,|\vec{p}_{\Lambda}|)\, ,
\qquad
q^{\mu}=(q_{0},0,0,-|\vec{q}\,|)\, ,
\end{align*}
where $q^{\mu}$ is the four-momentum of the virtual vector boson in the $\Lambda_{b}$ rest frame, and
\begin{align}
q_0 =&\frac{1}{2m_{\Lambda_{b}}}(m_{\Lambda_{b}}^2-m_{\Lambda}^2+q^2)\, ,
&E_{\Lambda}=&\frac{1}{2m_{\Lambda_{b}}}(m_{\Lambda_{b}}^2+m_{\Lambda}^2-q^2)\, ,\nn
\\
|\vec{q}\,| =&|\vec{p}_{\Lambda}|=\frac{1}{2m_{\Lambda_{b}}}\sqrt{Q_+Q_-}\, ,
&Q_\pm =& (m_{\Lambda_{b}} \pm m_{\Lambda})^2 - q^2\, .
\end{align}

The Dirac spinors  in Dirac representation can be written as
\begin{align}
u_{\Lambda_b}(\vec{p},\lambda_{\Lambda_b})=\sqrt{2m_{\Lambda_b}}
\begin{spmatrix}
\chi(\vec{p},\lambda_{\Lambda_b})
\\
0
\end{spmatrix},
\,\,
u_{\Lambda}(\vec{k},\lambda_{\Lambda})=
\begin{spmatrix}
\hphantom{2\lambda_{\Lambda}}\sqrt{E_{\Lambda}+m_{\Lambda}}\chi(\vec{k},\lambda_{\Lambda})
\\
2\lambda_{\Lambda}\sqrt{E_{\Lambda}-m_{\Lambda}}\chi(\vec{k},\lambda_{\Lambda})
\end{spmatrix}\, ,
\end{align}
where  $\chi(\vec{p},1/2)=\chi(\vec{k},1/2)=(1,0)^T,~\chi(\vec{p},-1/2)=\chi(\vec{k},-1/2)=(0,1)^T$\, . 

The polarization vectors for the virtual vector boson in the $\Lambda_{b}$ rest frame can be written as
\begin{align}\label{polvec}
\epsilon^{\mu}(t) =& \frac{1}{\sqrt{q^2}}\left(q_0,0,0,-|\vec{q}\,|\right)\, ,
\cr
\epsilon^{\mu}(0) =& \frac{1}{\sqrt{q^2}}\left(|\vec{q}\,|,0,0,-q_0\right)\,  ,
\cr
\epsilon^{\mu}(\pm) =& \frac{1}{\sqrt{2}}\left(0,\mp 1,i,0\right)\, ,
\end{align}
which satisfy the orthonormality and completeness relation Eq.~(\ref{eq:orthonormality and completeness}).

In the calculation of the leptonic helicity amplitudes, we work in the rest frame of the virtual vector boson, which is equivalent to the rest frame of the $\ell^+$-$\ell^-$ system. We have
\begin{align}
q^{\mu}=&(\sqrt{q^2},0,0,0)\, , \cr
p^\mu_{\ell^-} =&(E_\ell ,\hphantom{-}|\vec{p}_\ell|\sin\theta_\ell,0,\hphantom{-}|\vec{p}_\ell|\cos\theta_\ell)\, ,\cr
p^\mu_{\ell^+} =& (E_\ell ,-|\vec{p}_\ell|\sin\theta_\ell,0,-|\vec{p}_\ell|\cos\theta_\ell)\, ,
\end{align}
where $|\vec{p}_\ell| = \sqrt{q^2} \beta_\ell/2$ , $E_\ell = \sqrt{q^2}/2$ , $\beta_\ell\equiv\sqrt{1-{4m_\ell^2}/{q^2}}$, and $\theta_\ell$ denotes the angle between the three-momenta of $\ell^-$ and $\Lambda$. Moreover, we define $\beta_{\ell+}\equiv\sqrt{1+{4m_\ell^2}/{q^2}}$ for convenience. 

The Dirac spinors for $\ell^-$ and $\ell^+$ in Dirac representation read
\begin{align}\label{eq:Dirac spinor}
u_{\ell^-}(\vec{p}_\ell,\lambda_{\ell^-}) =&
\begin{spmatrix}
\hphantom{2\lambda_{\ell^-}}\sqrt{E_\ell + m_\ell}\chi(\vec{p}_\ell,\lambda_{\ell^-})
\\
2\lambda_{\ell^-}\sqrt{E_\ell - m_\ell}\chi(\vec{p}_\ell,\lambda_{{\ell^-}})
\end{spmatrix}\, ,\cr
v_{\ell^+}(-\vec{p}_\ell,\lambda_{\ell^+}) =& 
\begin{spmatrix}
\hphantom{-2\lambda_{\ell^+}}\sqrt{E_\ell - m_\ell} \xi(-\vec{p}_\ell,\lambda_{\ell^+})
\\
-2\lambda_{\ell^+}\sqrt{E_\ell + m_\ell}\xi(-\vec{p}_\ell,\lambda_{\ell^+})
\end{spmatrix}\, ,
\end{align}
respectively, which is consistent with the Jacob-Wick conventions in Ref.~\cite{Haber:1994pe}. $\chi(\vec{p}_\ell,\lambda_{\ell^-})$ and $\xi(-\vec{p}_\ell,\lambda_{\ell^+})$ read \cite{Yan:2019hpm},
\begin{align}
\xi(-\vec{p}_\ell,\frac{1}{2})=\chi(\vec{p}_\ell,\frac{1}{2})=&
\begin{spmatrix}
\cos\frac{\theta_\ell}{2} \\ \sin\frac{\theta_\ell}{2}
\end{spmatrix}\,,
&-\xi(-\vec{p}_\ell,-\frac{1}{2})=\chi(\vec{p}_\ell,-\frac{1}{2})=&
\begin{spmatrix}
-\sin\frac{\theta_\ell}{2} \\  \hphantom{-} \cos\frac{\theta_\ell}{2}
\end{spmatrix}\,.
\end{align}

The polarization vectors of the virtual vector boson in the bilepton rest frame are written as
\begin{align}\label{polvec2}
\bar{\epsilon}^{\mu}(t) = (1,0,0,0)\, ,
\qquad
\bar{\epsilon}^{\mu}(0) = (0,0,0,-1)\, ,
\qquad
\bar{\epsilon}^{\mu}(\pm) = \frac{1}{\sqrt{2}}(0,\mp 1,i,0)\, ,
\end{align}
which can also be obtained from Eq.~(\ref{polvec}) by a Lorentz transformation and satisfy the orthonormality and completeness relation in Eq.~(\ref{eq:orthonormality and completeness}).

The phase space can be generated recursively~\cite{PDG:2018}:
\begin{align}
{\rm d}\Phi_n(p;p_1,p_2,\cdots,p_n)=\int {\rm d}\Phi_{j}(q;p_1,\cdots,p_j){\rm d}\Phi_{n-j+1}(p;q,p_{j+1},\cdots,p_n)\frac{{\rm d}q^2}{2\pi}\, .
\end{align}	
The four-body phase space can be decomposed into three two-body as follows, 
\begin{align}
{\rm d}\Phi_4(p;k_1,k_2,q_1,q_2)=&(2\pi)^4\delta^4(p-k_1-k_2-q_1-q_2)\prod_{i=1}^{2}\frac{{\rm d}^3\vec{k}_i}{(2\pi)^32E_{k_i}}\prod_{i=1}^{2}\frac{{\rm d}^3\vec{q}_i}{(2\pi)^32E_{q_i}}\nn\\
=&\int\frac{{\rm d}q^2}{2\pi}\frac{{\rm d}k^2}{2\pi}{\rm d}\Phi_2(k;k_1,k_2){\rm d}\Phi_2(q;q_1,q_2){\rm d}\Phi_2(p;k,q)\, ,
\end{align}
where the corresponding two-body phase space can be written as 
\begin{align}
\int {\rm d}\Phi_2(k;k_1,k_2)=&\frac{1}{2^5\pi^2}\frac{\sqrt{\lambda(k^2,k_1^2,k_2^2)}}{k^2}\int_{-1}^{1}{\rm d}\cos\theta_\Lambda\int_{0}^{2\pi}{\rm d}\phi \,,\nn\\
\int {\rm d}\Phi_2(q;q_1,q_2)=&\frac{1}{2^5\pi^2}\frac{\sqrt{\lambda(q^2,q_1^2,q_2^2)}}{q^2}\int_{-1}^{1}{\rm d}\cos\theta_\ell\times(2\pi)\, ,\nn\\
\int {\rm d}\Phi_2(p;k,q)=&\frac{1}{2^5\pi^2}\frac{\sqrt{\lambda(p^2,k^2,q^2)}}{p^2}\times 2\times(2\pi) \,.
\end{align}
In summary, we can get the four body phase space formula as follows, 
\begin{align}\label{eq:fourbody}
\int {\rm d}\Phi_4(p;k_1,k_2,q_1,q_2)=&\frac{1}{2^{14}\pi^6}\int \frac{1}{p^2 q^2}{\rm d}q^2{\rm d}\cos\theta_\ell{\rm d}\cos\theta_\Lambda {\rm d}\phi \sqrt{\lambda(p^2,k^2,q^2)} \nn\\ 
&\times\sqrt{\lambda(q^2,q_1^2,q_2^2)}\frac{1}{k^2}\sqrt{\lambda(k^2,k_1^2,k_2^2)}~{\rm d}k^2\nn\\
\equiv&\int {\rm d}\bar{\Phi}_4(p;k_1,k_2,q_1,q_2) {\rm d}k^2\,.
\end{align}

\section{Helicity amplitudes}\label{sec:helicity}
\subsection{Hadronic helicity amplitudes}
The hadronic helicity amplitudes $N_1\to N_2$ are defined as
\begin{align}\label{eq:def H}
  H^{\rm S}_{\lambda_{N_1},\lambda_{N_2}}=&\bra{N_2(\lambda_{N_2})}\bar{s}\, b\ket{N_1(\lambda_{N_1})}\,,\nonumber\\
  H^{\rm P}_{\lambda_{N_1},\lambda_{N_2}}=&\bra{N_2(\lambda_{N_2})}\bar{s}\,\gamma_5\, b\ket{N_1(\lambda_{N_1})}\,,\nonumber\\  
  H^{\rm V}_{\lambda_{N_1},\lambda_{N_2},\lambda_{W}}=&\,\epsilon^{\mu*}(\lambda_{W})\bra{N_2(\lambda_{N_2})}\bar{s}\,\gamma_{\mu}\, b\ket{N_1(\lambda_{N_1})}\, , \nonumber\\ 
  H^{\rm A}_{\lambda_{N_1},\lambda_{N_2},\lambda_{W}}=&\,\epsilon^{\mu*}(\lambda_{W})\bra{N_2(\lambda_{N_2})}\bar{s}\,\gamma_{\mu}\gamma_5\, b\ket{N_1(\lambda_{N_1})}\,,\nonumber\\
  H^{{\rm Tq},\lambda_{N_1}}_{\lambda_{N_2},\lambda_{W1} }=&\epsilon^{\mu*}(\lambda_{W1})\bra{N_2(\lambda_{N_2})}\bar{s}\,\tred{i}\sigma_{\mu \nu}q^{\nu} \,b\ket{N_1(\lambda_{N_1})}\,, \nonumber\\
H^{{\rm Tq5},\lambda_{N_1}}_{\lambda_{N_2},\lambda_{W1} }=&\epsilon^{\mu*}(\lambda_{W1})\bra{N_2(\lambda_{N_2})}\bar{s}\,\tred{i}\sigma_{\mu \nu}q^{\nu}\gamma_5 \,b\ket{N_1(\lambda_{N_1})}\, ,\nonumber\\
  H^{{\rm T0},\lambda_{N_1}}_{\lambda_{N_2},\lambda_{W1} ,\lambda_{W2}}=&\tblue{i}\epsilon^{\mu*}(\lambda_{W1})\epsilon^{\nu*}(\lambda_{W2})\bra{N_2(\lambda_{N_2})}\bar{s}\,\sigma_{\mu \nu}\, b\ket{N_1(\lambda_{N_1})}\,, \nonumber\\
H^{{\rm T5},\lambda_{N_1}}_{\lambda_{N_2},\lambda_{W1} ,\lambda_{W2}}=&\tblue{i}\epsilon^{\mu*}(\lambda_{W1})\epsilon^{\nu*}(\lambda_{W2})\bra{N_2(\lambda_{N_2})}\bar{s}\,\sigma_{\mu \nu}\gamma_5\, b\ket{N_1(\lambda_{N_1})}\,,
\end{align}
where~$q^\mu=p_M^\mu-p_N^\mu$,~$\epsilon^{\mu}$ are the polarization vectors of virtual vector boson in the $N_1$ rest frame~\cite{Yan:2019hpm}. It is straightforward to obtain $H^{T0,\lambda_{M}}_{\lambda_{N},\lambda_{W1}\,  ,\lambda_{W2}}=-H^{T0,\lambda_{M}}_{\lambda_{N},\lambda_{W2} ,\lambda_{W1}}$.
\subsection{Leptonic helicity amplitudes}
The leptonic helicity amplitudes  are defined as ~\cite{Yan:2019hpm}
\begin{align}\label{eq:Ldef}
  L^{\rm S}_{\lambda_{\ell_1},\lambda_{\ell_2}}=&\bra{\ell_1\ell_2}\bar\ell_1\ell_2\ket{0}=\bar{u}_{\ell_1}(\vec{p}_{\ell_1},\lambda_{\ell_1})v_{\ell_2}(-\vec{p}_{\ell_1},\lambda_{\ell_2})\, ,
  \nn\\
   L^{\rm P}_{\lambda_{\ell_1},\lambda_{\ell_2}}=&\bra{\ell_1\ell_2}\bar\ell_1\gamma_5\ell_2\ket{0}=\bar{u}_{\ell_1}(\vec{p}_{\ell_1},\lambda_{\ell_1})\gamma_5 v_{\ell_2}(-\vec{p}_{\ell_1},\lambda_{\ell_2})\, ,
  \nn\\
  L^{\rm V}_{\lambda_{\ell_1},\lambda_{\ell_2},\lambda_W}=&\bar{\epsilon}^{\mu} (\lambda_W)\bra{\ell_1\ell_2}\bar\ell_1\gamma_\mu \ell_2\ket{0} =\bar{\epsilon}^{\mu}(\lambda_W)\bar{u}_{\ell_1}(\vec{p}_{\ell_1},\lambda_{\ell_1})\gamma_\mu v_{\ell_2}(-\vec{p}_{\ell_1},\lambda_{\ell_2}) \, , \nn\\
   L^{\rm A}_{\lambda_{\ell_1},\lambda_{\ell_2},\lambda_W}=&\bar{\epsilon}^{\mu} (\lambda_W)\bra{\ell_1\ell_2}\bar\ell_1\gamma_\mu \gamma_5 \ell_2\ket{0} =\bar{\epsilon}^{\mu}(\lambda_W)\bar{u}_{\ell_1}(\vec{p}_{\ell_1},\lambda_{\ell_1})\gamma_\mu \gamma_5 v_{\ell_2}(-\vec{p}_{\ell_1},\lambda_{\ell_2}) \, , \nn\\
  L^{\rm T0}_{\lambda_{\ell_1},\lambda_{\ell_2},\lambda_{W_1} ,\lambda_{W_2}} =&-i\bar{\epsilon}^{\mu} (\lambda_{W_1})\bar{\epsilon}^{\nu} (\lambda_{W_2})\bra{\ell_1\ell_2}\bar\ell_1\sigma_{\mu \nu} \ell_2\ket{0} \nn\\
  =&-i\bar{\epsilon}^{\mu} (\lambda_{W_1})\bar{\epsilon}^{\nu} (\lambda_{W_2})\bar{u}_{\ell_1}(\vec{p}_{\ell_1},\lambda_{\ell_1})\sigma_{\mu \nu} v_{\ell_2}(-\vec{p}_{\ell_1},\lambda_{\ell_2})\, ,\nn\\
  L^{\rm T5}_{\lambda_{\ell_1},\lambda_{\ell_2},\lambda_{W_1} ,\lambda_{W_2}} =&-i\bar{\epsilon}^{\mu} (\lambda_{W_1})\bar{\epsilon}^{\nu} (\lambda_{W_2})\bra{\ell_1\ell_2}\bar\ell_1\sigma_{\mu \nu} \gamma_5 \ell_2\ket{0} \nn\\
  =&-i\bar{\epsilon}^{\mu} (\lambda_{W_1})\bar{\epsilon}^{\nu} (\lambda_{W_2})\bar{u}_{\ell_1}(\vec{p}_{\ell_1},\lambda_{\ell_1})\sigma_{\mu \nu} \gamma_5 v_{\ell_2}(-\vec{p}_{\ell_1},\lambda_{\ell_2})\, ,
\end{align}
where $\bar{\epsilon}^{\mu}$~are the polarization vectors in the virtual vector boson rest frame~\cite{Yan:2019hpm}.

According to the definition of lepton helicity amplitude, the non-zero results of vector and scalar read
\begin{align}\label{L:b2s1}
  L^{\rm S}_{\pm 1/2,\pm 1/2}=& \pm\sqrt{q^2}\beta_\ell\,,
  &L^{\rm P}_{\pm 1/2,\pm 1/2}=&-\sqrt{q^2}\,,
  \nn\\
  L^{\rm V}_{1/2,1/2,\pm}=&L^{\rm V}_{-1/2,-1/2,\mp}=\mp\sqrt{2}m_\ell \sin\theta_\ell \,,
  &L^{\rm V}_{\pm 1/2,\pm 1/2,0}=&\mp 2m_\ell \cos\theta_\ell\,,
  \nn\\
  L^{\rm V}_{\pm 1/2,\mp 1/2,\pm}=&\pm\frac{\sqrt{q^2}}{\sqrt{2}}(1 -\cos\theta_\ell)\,,
  &L^{\rm V}_{\pm 1/2,\mp 1/2,0}=&\sqrt{q^2}\sin\theta_\ell\,,
  \nn\\
  L^{\rm V}_{\pm 1/2,\mp 1/2,\mp}=&\pm\frac{\sqrt{q^2}}{\sqrt{2}}(1+\cos\theta_\ell)\,,
  &L^{\rm A}_{\pm 1/2,\pm 1/2,t}=&-2m_\ell\,,\nn\\
  L^{\rm A}_{\pm 1/2,\mp 1/2,\pm}=&\frac{\sqrt{q^2}}{\sqrt{2}}(1 -\cos\theta_\ell)\beta_\ell\,,
  &L^{\rm A}_{\pm 1/2,\mp 1/2,0}=&\pm\sqrt{q^2}\sin\theta_\ell\beta_\ell\,,\nn\\
  L^{\rm A}_{\pm 1/2,\mp 1/2,\mp}=&\frac{\sqrt{q^2}}{\sqrt{2}}(1 +\cos\theta_\ell)\beta_\ell\,.
\end{align}
The non-zero tensor amplitudes are
\begin{align}\label{L:b2s2}
  L^{{\rm T0},1/2}_{1/2,t,\pm}=&-L^{{\rm T0},-1/2}_{-1/2,t,\pm}=\mp L^{{\rm T5},1/2}_{1/2,\pm,0}=\pm L^{{\rm T5},-1/2}_{-1/2,\pm,0}=\mp\frac{\sqrt{q^2} \sin\theta_\ell}{\sqrt{2}}\,, \nn\\
  L^{{\rm T0},1/2}_{1/2,\pm,0}=&L^{{\rm T0},-1/2}_{-1/2,\pm,0}=\mp L^{{\rm T5},1/2}_{1/2,t,\pm}=\mp L^{{\rm T5},-1/2}_{-1/2,t,\pm}=-\frac{\beta_\ell  \sqrt{q^2} \sin \theta_\ell}{\sqrt{2}}\,,\nn\\
  L^{{\rm T0},\pm 1/2}_{\mp 1/2,t,\pm}=&\mp L^{{\rm T5},\pm 1/2}_{\mp 1/2,\pm,0}=\pm\sqrt{2} (1-\cos \theta_\ell) m_{\ell }\,,\nn\\
  L^{{\rm T0},\pm 1/2}_{\mp 1/2,t,\mp}=&\pm L^{{\rm T5},\pm 1/2}_{\mp 1/2,\mp,0}=\pm\sqrt{2} (1+\cos \theta_\ell ) m_{\ell }\,,\nn\\
  L^{{\rm T0},\pm 1/2}_{\pm 1/2,+,-}=&-L^{{\rm T5},\pm 1/2}_{\pm 1/2,t,0}=-\sqrt{q^2}\beta_\ell \cos\theta_\ell\,,\nn\\
  L^{{\rm T0},\pm 1/2}_{\pm 1/2,t,0}=&L^{{\rm T5},\mp 1/2}_{\mp 1/2,+,-}=\mp\sqrt{q^2} \cos\theta_\ell\,,\nn\\
  L^{{\rm T0},\pm 1/2}_{\mp 1/2,t,0}=&L^{{\rm T5},\pm 1/2}_{\mp 1/2,-,+}=2\sin\theta_\ell m_{\ell }\,.
\end{align}

\section{Determination of the angular distribution \label{sec:4body}}
In this appendix, we introduce a method called narrow-width approximation to calculate the angular distribution of four-body processes \cite{Kim:2000dq,Altmannshofer:2008dz,Descotes-Genon:2019dbw}. We take $\Lambda_b\to \Lambda(\to N\pi)\ell^+\ell^-$ decay as an example to illustrate this method.

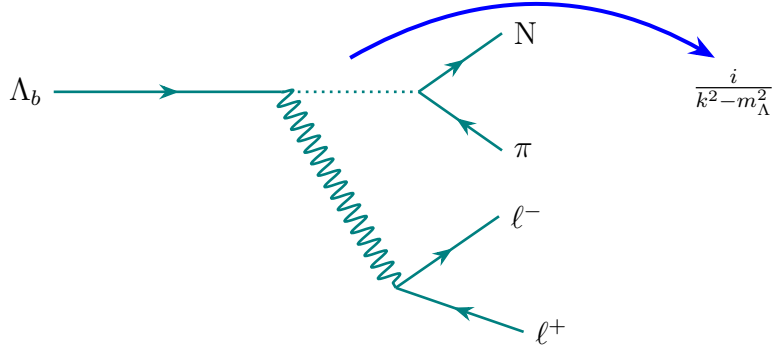
\begin{figure}
\begin{center}
\begin{tikzpicture}[line width=1pt,scale=1.5,>=Stealth]
	\begin{scope}
	\path (0,0) coordinate (a0);
	\path (1.2,0) coordinate (a1);
	\path (300:2) coordinate (a2);
	\path (180:2) coordinate (a3);
	\path (15:2) coordinate (a4);
	\path (345:2) coordinate (a5);
	\path (330:2.2) coordinate (a6);
	\path (315:3) coordinate (a7);
	\draw [scalarnoarrow](a0)--(a1);
	\draw [photon](a0)--(a2);
	\draw [fermion](a3)--(a0);
	\draw [fermion](a1)--(a4);
	\draw [fermion](a5)--(a1);
	\draw [fermion](a2)--(a6);
	\draw [fermion](a7)--(a2);
	\draw [->,blue,opacity=1,overlay,line width=1.5pt](0.6,0.3) to [bend left] (3.8,0.3);
	\node [left] at  (a3) {$\Lambda_b$};
	\node [right] at  (a4) {\text{N}};
	\node [right] at  (a5) {$\pi$};
	\node [right] at  (a6) {$\ell^-$};
	\node [right] at  (a7) {$\ell^+$};
	\node [right] at  (3.5,0) {$\frac{i}{k^2-m_\Lambda^2}$};
	\end{scope}
	\end{tikzpicture}
	\caption{\label{fig:feyndiag} Feynman diagram of the decay $\Lambda_{b}\to\Lambda(\to N\pi)\ell^+\ell^-$.}
\end{center}
\end{figure}

The conventions are collected as follows
\begin{align}
H(s_{\Lambda_b},s_{\Lambda})\equiv & ~\bra{\Lambda(s_{\Lambda})}\mathcal{O}\ket{\Lambda_b(s_{\Lambda_b})}\,,\nn\\
H(s_{\Lambda_b},s_{\Lambda},\lambda)\equiv & ~\epsilon^{\mu*}(\lambda)\bra{\Lambda(s_{\Lambda})}\mathcal{O}_\mu\ket{\Lambda_b(s_{\Lambda_b})}\,,\nn\\
H(s_{\Lambda_b},s_{\Lambda},\lambda_1,\lambda_2)\equiv & ~\epsilon_1^{\mu*}(\lambda_1)\epsilon_2^{\nu*}(\lambda_2) \bra{\Lambda(s_{\Lambda})}\mathcal{O}_{\mu\nu}\ket{\Lambda_b(s_{\Lambda_b})}\,,\nn\\
L(s_{\ell^+},s_{\ell^-})\equiv & ~\bra{\ell^-(s_{\ell^-})}\mathcal{O}\ket{\ell^+(s_{\ell^+})}\,,\nn\\
L(s_{\ell^+},s_{\ell^-},\lambda)\equiv & ~\bar{\epsilon}^{\mu}(\lambda)\bra{\ell^-(s_{\ell^-})}\mathcal{O}_\mu\ket{\ell^+(s_{\ell^+})}\,,\nn\\
L(s_{\ell^+},s_{\ell^-},\lambda_1,\lambda_2)\equiv & ~\bar{\epsilon}_1^{\mu}(\lambda_1)\bar{\epsilon}_2^{\nu}(\lambda_2) \bra{\ell^-(s_{\ell^-})}\mathcal{O}_{\mu\nu}\ket{\ell^+(s_{\ell^+})}\,.
\end{align}
They correspond to the helicity amplitudes of scalar (pseudoscalar), vector (axial vector) and tensor operators, respectively.

~$\Lambda\to N\pi$~amplitude can be written as~$H(s_{\Lambda},s_{N})\equiv\bra{N(s_{N})\pi}\mathcal{O}\ket{\Lambda(s_{\Lambda})}$. The $\Lambda$ propagator is recorded as $\frac{i}{k^2-m_{\Lambda}^2}$, as shown in Fig. \ref{fig:feyndiag}. Based on the above discussion, the invariant amplitude of $\Lambda_b\to \Lambda(N\pi)\ell^+\ell^-$ reads \footnote{The coefficient  $\tred{\frac{1}{2}}$~comes from $P_{\rm L(R)}$ in ~$b\to s\ell^+\ell^-$~ effective Hamiltonian.}
\begin{align}
\mathcal{M}_0(s_{\Lambda_b},s_{N},s_{\ell^+},s_{\ell^-})=&~\bra{N(s_{N})\pi\ell^+(s_{\ell^+})\ell^-(s_{\ell^-})}\mathcal{H}_{eff} \ket{\Lambda_b(s_{\Lambda_b})}\nn\\
=&~\sum_{s_{\Lambda}}\bra{\Lambda(s_{\Lambda})\pi\ell^+(s_{\ell^+})\ell^-(s_{\ell^-})}\mathcal{H}_{eff} \ket{\Lambda_b(s_{\Lambda_b})}\nn\\
&\times \frac{i}{k^2-m_{\Lambda}^2} \bra{N(s_{N})\pi}\mathcal{O}\ket{\Lambda(s_{\Lambda})}\nn\\
=&\frac{i}{k^2-m_{\Lambda}^2}\sum_{s_{\Lambda}}H(s_{\Lambda},s_{N})\big[\tred{\frac{1}{2}} H(s_{\Lambda_b},s_{\Lambda})L(s_{\ell^+},s_{\ell^-})\nn\\
&+\tred{\frac{1}{2}}\sum_{\lambda}g_{\lambda,\lambda}H(s_{\Lambda_b},s_{\Lambda},\lambda)L(s_{\ell^+},s_{\ell^-},\lambda)\nn\\
&+\sum_{\lambda_1,\lambda_2}g_{\lambda_1,\lambda_1}g_{\lambda_2,\lambda_2}H(s_{\Lambda_b},s_{\Lambda},\lambda_1,\lambda_2)L(s_{\ell^+},s_{\ell^-},\lambda_1,\lambda_2)\big]\nn\\
\equiv & \frac{i}{k^2-m_{\Lambda}^2}~\mathcal{M}_1(s_{\Lambda_b},s_{N},s_{\ell^+},s_{\ell^-})\,.
\end{align}

After averaging over the helicity of $\Lambda_b$ , we get
\begin{align}
\abs{\mathcal{M}}^2=&\frac{1}{2}\sum_{s_{\Lambda_b}}\sum_{s_{N}}\sum_{s_{\ell^+}}\sum_{s_{\ell^-}}\abs{\mathcal{M}_0(s_{\Lambda_b},s_{N},s_{\ell^+},s_{\ell^-})}^2\cr
=&~\frac{1}{(k^2-m_{\Lambda}^2)^2}{\frac{1}{2}}\sum_{s_{\Lambda_b}}\sum_{s_{N}}\sum_{s_{\ell^+}}\sum_{s_{\ell^-}}\abs{\mathcal{M}_1(s_{\Lambda_b},s_{N},s_{\ell^+},s_{\ell^-})}^2 \cr
\equiv& \frac{\abs{N}^2}{(k^2-m_{\Lambda}^2)^2}{\frac{1}{2}}\,,
\end{align}
\begin{align}
\abs{N}^2\equiv&\sum_{s_{\Lambda_b}}\sum_{s_{N}}\sum_{s_{\ell^+}}\sum_{s_{\ell^-}}\abs{\mathcal{M}_1(s_{\Lambda_b},s_{N},s_{\ell^+},s_{\ell^-})}^2\nn\\
=&\sum_{s_{\Lambda_b}}\sum_{s_{\ell^+}}\sum_{s_{\ell^-}}\sum_{s_{\Lambda}}\sum_{s_{\Lambda}^{\prime}}~[\sum_{s_{N}}H(s_{\Lambda},s_{N})H^{*}(s_{\Lambda}^{\prime},s_{N})]\nn\\
&\times\Big[\tred{\frac{1}{4}}H(s_{\Lambda_b},s_{\Lambda})L(s_{\ell^+},s_{\ell^-})H^{*}(s_{\Lambda_b},s_{\Lambda}^{\prime})L^{*}(s_{\ell^+},s_{\ell^-})\nn\\
&+\tred{\frac{1}{4}}\sum_{\lambda}\sum_{\lambda^{\prime}}g_{\lambda,\lambda}g_{\lambda^{\prime}\lambda^{\prime}}H(s_{\Lambda_b},s_{\Lambda},\lambda)L(s_{\ell^+},s_{\ell^-},\lambda)H^{*}(s_{\Lambda_b},s_{\Lambda}^{\prime},\lambda^{\prime})L^{*}(s_{\ell^+},s_{\ell^-},\lambda^{\prime})\nn\\
&+\sum_{\lambda_1,\lambda_2}\sum_{\lambda_1^{\prime},\lambda_2^{\prime}}g_{\lambda_1,\lambda_1}g_{\lambda_2,\lambda_2}g_{\lambda_1^{\prime},\lambda_1^{\prime}}g_{\lambda_2^{\prime},\lambda_2^{\prime}}H(s_{\Lambda_b},s_{\Lambda},\lambda_1,\lambda_2)L(s_{\ell^+},s_{\ell^-},\lambda_1,\lambda_2)\nn\\
&\times H^{*}(s_{\Lambda_b},s_{\Lambda}^{\prime},\lambda_1^{\prime},\lambda_2^{\prime})L^{*}(s_{\ell^+},s_{\ell^-},\lambda_1^{\prime},\lambda_2^{\prime})\nn\\
&+\tred{\frac{1}{4}}\sum_{\lambda}g_{\lambda,\lambda}[H(s_{\Lambda_b},s_{\Lambda},\lambda)L(s_{\ell^+},s_{\ell^-},\lambda)H^{*}(s_{\Lambda_b},s_{\Lambda}^{\prime})L^{*}(s_{\ell^+},s_{\ell^-})\nn\\
&+H^{*}(s_{\Lambda_b},s_{\Lambda}^{\prime},\lambda)L^{*}(s_{\ell^+},s_{\ell^-},\lambda)H(s_{\Lambda_b},s_{\Lambda})L(s_{\ell^+},s_{\ell^-})]\nn\\
&+\tred{\frac{1}{2}}\sum_{\lambda_1,\lambda_2}g_{\lambda_1,\lambda_1}g_{\lambda_2,\lambda_2}[H(s_{\Lambda_b},s_{\Lambda},\lambda_1,\lambda_2)L(s_{\ell^+},s_{\ell^-},\lambda_1,\lambda_2)H^{*}(s_{\Lambda_b},s_{\Lambda}^{\prime})L^{*}(s_{\ell^+},s_{\ell^-})\nn\\
&+H^{*}(s_{\Lambda_b},s_{\Lambda}^{\prime},\lambda_1,\lambda_2)L^{*}(s_{\ell^+},s_{\ell^-},\lambda_1,\lambda_2)H(s_{\Lambda_b},s_{\Lambda})L(s_{\ell^+},s_{\ell^-})]\nn\\
&+\tred{\frac{1}{2}}\sum_{\lambda_1,\lambda_2}\sum_{\lambda^{\prime}}g_{\lambda_1,\lambda_1}g_{\lambda_2,\lambda_2}g_{\lambda^{\prime},\lambda^{\prime}}\nn\\
&\times [H(s_{\Lambda_b},s_{\Lambda},\lambda_1,\lambda_2)L(s_{\ell^+},s_{\ell^-},\lambda_1,\lambda_2)H^{*}(s_{\Lambda_b},s_{\Lambda}^{\prime},\lambda^{\prime})L^{*}(s_{\ell^+},s_{\ell^-},\lambda^{\prime})\nn\\
&+H^{*}(s_{\Lambda_b},s_{\Lambda}^{\prime},\lambda_1,\lambda_2)L^{*}(s_{\ell^+},s_{\ell^-},\lambda_1,\lambda_2)H(s_{\Lambda_b},s_{\Lambda},\lambda^{\prime})L(s_{\ell^+},s_{\ell^-},\lambda^{\prime})]\Big]\,.
\end{align}

By Eq.~(\ref{eq:fourbody}), we can get ~\cite{Altmannshofer:2008dz,Descotes-Genon:2019dbw}
\begin{align}
\int {\rm d}\Phi_4\frac{\abs{M}^2}{2m_{\Lambda_b}}=&\int {\rm d}\bar{\Phi}_4 {\rm d}k^2\frac{1}{2m_{\Lambda_b}}\frac{\abs{N}^2}{(k^2-m_{\Lambda}^2)^2}\tblue{\frac{1}{2}}\nn\\
\xrightarrow{\Gamma_{\Lambda}\ll m_{\Lambda}} & \int {\rm d}\bar{\Phi}_4 {\rm d}k^2\frac{\abs{N}^2}{2^2m_{\Lambda_b}}\frac{1}{(k^2-m_{\Lambda}^2)^2+(m_{\Lambda}\Gamma_{\Lambda})^2}\nn\\
\rightarrow & \frac{1}{2^2m_{\Lambda_b}}\int {\rm d}\bar{\Phi}_4 {\rm d}k^2 \abs{N}^2\frac{\pi}{m_{\Lambda}\Gamma_{\Lambda}}\delta(k^2-m_{\Lambda}^2)\nn\\
=& \frac{1}{2^2m_{\Lambda_b}}\frac{\pi}{m_{\Lambda}\Gamma_{\Lambda}}\int {\rm d}\bar{\Phi}_4  \abs{N}^2|_{k^2=m_{\Lambda}^2}\nn\\
=&\frac{1}{2^{12}\pi^4 m_{\Lambda_b}^3}\int \frac{1}{q^2}{\rm d}q^2{\rm d}\cos\theta_\ell{\rm d}\cos\theta_\Lambda {\rm d}\phi \sqrt{\lambda(m_{\Lambda_b}^2,m_{\Lambda}^2,q^2)} \sqrt{\lambda(q^2,m_{\ell^+}^2,m_{\ell^-}^2)}\nn\\ 
&\times\frac{1}{\Gamma_{\Lambda}}\frac{1}{2^4\pi m_{\Lambda}^3}\sqrt{\lambda(m_{\Lambda}^2,m_N^2,m_{\pi}^2)}~\abs{N}^2|_{k^2=m_{\Lambda}^2}\,.
\end{align}

A brief proof of narrow width approximate formula is as follows. 
Dirac delta function has the following two properties 
\begin{align}
\lim_{\epsilon \to 0}\frac{\epsilon}{\epsilon^2+x^2}=\pi\delta (x)\,,\qquad
\delta(\frac{k^2}{m_{\Lambda}^2}-1)=
m_\Lambda^2\delta(k^2-m_\Lambda^2)\,.
\end{align}
Thus, we can obtain
\begin{align}
\frac{1}{(k^2-m_{\Lambda}^2)^2+(m_{\Lambda}\Gamma_{\Lambda})^2}=&\frac{1}{m_{\Lambda}^3\Gamma_{\Lambda}}\frac{\frac{\Gamma_{\Lambda}}{m_{\Lambda}}}{(\frac{\Gamma_{\Lambda}}{m_{\Lambda}})^2+(\frac{k^2}{m_\Lambda^2}-1)^2}\nn\\
\xrightarrow{\Gamma_{\Lambda}\ll m_{\Lambda}} &\frac{1}{m_{\Lambda}^3\Gamma_{\Lambda}}\pi \delta(\frac{k^2}{m_\Lambda^2}-1)\nn\\
=&\frac{\pi}{m_{\Lambda}\Gamma_{\Lambda}}\delta(k^2-m_\Lambda^2)\,.
\end{align}

So far, we have deduced all the formulas involved to calculate four body angular distribution.
\end{appendix}

\bibliographystyle{JHEP}
\bibliography{ref}

\end{document}